\documentclass[times]{qjrms4}

\usepackage[colorlinks,bookmarksopen,bookmarksnumbered,citecolor=red,urlcolor=red]{hyperref}
\usepackage{natbib}
\usepackage{caption}
\usepackage{tabularx}
\newcommand\BibTeX{{\rmfamily B\kern-.05em \textsc{i\kern-.025em b}\kern-.08em
T\kern-.1667em\lower.7ex\hbox{E}\kern-.125emX}}

\usepackage{moreverb, amsmath}

\def\volumeyear{2019}

\begin{document}

\runningheads{K. Strommen}{Jet latitude regimes and NAO predictability}

\title{Jet Latitude Regimes and the Predictability of the North Atlantic Oscillation}

\author{K.~Strommen\corrauth}

\address{Oxford University, Clarendon Laboratory, AOPP, Parks Road, OX1 3PU, Oxford.}

\corraddr{Oxford University, Clarendon Laboratory, AOPP, Parks Road, OX1 3PU, Oxford. E-mail: kristian.strommen@physics.ox.ac.uk}

\begin{abstract}
In recent years, numerical weather prediction models have begun to show notable levels of skill at predicting the average winter North Atlantic Oscillation (NAO) when initialised one month ahead. At the same time, these model predictions exhibit unusually low signal-to-noise ratios, in what has been dubbed a `signal-to-noise paradox'. We analyse both the skill and signal-to-noise ratio of the Integrated Forecast System (IFS), the European Center for Medium-range Weather Forecasts (ECMWF) model, in an ensemble hindcast experiment. Specifically, we examine the contribution to both from the regime dynamics of the North Atlantic eddy-driven jet. This is done by constructing a statistical model which captures the predictability inherent to to the trimodal jet latitude system, and fitting its parameters to reanalysis and IFS data. Predictability in this regime system is driven by interannual variations in the persistence of the jet latitude regimes, which determine the preferred state of the jet. We show that the IFS has skill at predicting such variations in persistence: because the position of the jet strongly influences the NAO, this automatically generates skill at predicting the NAO. We show that all of the skill the IFS has at predicting the winter NAO over the period 1980-2010 can be attributed to its skill at predicting regime persistence in this way. Similarly, the tendency of the IFS to underestimate regime persistence can account for the low signal-to-noise ratio, giving a possible explanation for the signal-to-noise paradox. Finally, we examine how external forcing drives variability in jet persistence, as well as highlight the role played by transient baroclinic eddy feedbacks to modulate regime persistence.
\end{abstract}


\maketitle

\section{Introduction} 
\label{sec:intro}

The leading mode of variability in the North Atlantic region is referred to as the North Atlantic Oscillation (NAO), which captures a frequently recurring dipole structure of pressure anomalies. During the winter season the NAO is to a large extent capturing the location of the storm track \citep{Rogers1997} and therefore has a big influence on European winter climate. Despite the fact that, due to the chaotic nature of the atmosphere, any individual storm is not predictable for more than a few weeks ahead, evidence has begun to accumulate which suggests that the \emph{average} winter conditions may be predictable \citep{Smith2016}. Furthermore, modern numerical weather prediction (NWP) models have begun to show notable and consistent levels of skill at capturing this predictable signal when initialized one month ahead (though it should be noted that comparable skill was already reported in \citet{Muller2005}). In a particularly noteworthy example, studies conducted by the UK Met Office system `GloSea5' \citep{Scaife2014a, Dunstone2016} showed that their seasonal prediction system displayed skilful ensemble hindcasts over the 35 year period 1980-2015. Subsequently, the multi-model study \citet{Athanasiadis2017} showed that comparable levels of skill was attainable by other models as well.
 
Two questions have naturally sprung out of all this work. Firstly, can one identify concrete external drivers that account for the majority of interannual NAO variability? A significant number of studies have attempted to address this question, both in terms of locating drivers of the observed NAO as well as studying which of these drivers are responsible for model skill. A good overview of these can be found in \citet{Smith2016}. Some of the more consistently proposed drivers include Atlantic sea surface temperatures \citep{Czaja1999, Rodwell1999, Saunders2002, Dunstone2016}, the El Ni{\~{n}}o Southern Oscillation (ENSO) \citep{Dong2000, Moron2003, Toniazzo2006, Bronnimann2007, Li2012, Drouard2013, Jimenez-Esteve2018} the stratospheric polar vortex \citep{Baldwin2001, Dunstone2016, Kidston2015}, the quasi biennial oscillation (QBO) \citep{Anstey2013, Scaife2014, OReilly2019} and Arctic sea ice, particularly in the Kara region \citep{Strong2011, Koenigk2016, Wang2017}.

The second big question arises from the observation, first noted in \citet{Eade2014} for GloSea5 and subsequently corroborated by the multi-model study \citet{Baker2018}, that the signal-to-noise ratio of model NAO ensemble predictions tends to be too low. In other words, the predictable component of the model appears to be a lot smaller than that of the real world. This phenomenon has been described as a `signal-to-noise paradox' in \citet{Dunstone2016}; see \citet{Scaife2018} for an overview. As a consequence, large ensembles are required in order for the ensemble mean to amplify the predictable signal sufficiently. Because a resolution of this `paradox' could lead to more skilful predictions, or equivalent skill with a smaller ensemble, it is of great interest to understand what is producing this low signal-to-noise ratio in models. 

In \citet{Kumar2014} and \citet{Siegert2016}, `linear regression style' statistical models aimed at representing NAO predictability were analysed, suggesting that the paradox may be due to the model not adequately representing a linear predictable signal in the \emph{magnitude} of the NAO. This implies that one possible explanation is one or more overly weak teleconnections in the model. Both North Atlantic SSTs \citep{Rodwell1999, Mehta2000, Gastineau2013, Scaife2013} and the QBO \citep{OReilly2019} have been suggested as examples of such.

An alternative hypothesis was proposed in \citet{Strommen2019}, based on the idea that the NAO is driven by regime dynamics. The possible influence of regimes on North Atlantic circulation has been studied since the 80's, often by way of classifying distinct, frequently recurring and quasi-persistent geopotential height patterns, of which the two phases of the NAO are examples \citep{Vautard1990, Michelangeli1995a, Dawson2012}. Studies have shown that both frequency of occurrence of individual regimes and the transition rates between different regimes can be influenced by external forcing \citep{Charlton-Perez2018}; it was suggested in \citet{Palmer1999} that anthropogenic forcing may also influence regime behaviour in a similar manner. In \citet{Strommen2019}, a statistical model was presented which captures the predictability inherent to such regime dynamics on seasonal time-scales. By envisaging the atmosphere as a two-state system, with transitions taking place on daily time-scales, it was shown that interannual variability in such a system is determined by changes in the preferred regime state each year and that this preference is in turn driven by variations in the persistence time-scales of the two regimes. Furthermore, if NWP models systematically underestimate the level of regime persistence relative to the real world, then the signal-to-noise paradox emerges as a natural feature of the system. That models struggle to replicate many aspects of regimes, including persistence, has been observed in e.g. \citet{Strommen2019a}, which highlighted the importance of horizontal resolution.

The purpose of the present paper is to build on the framework of \citet{Strommen2019} in three key ways. Firstly, the statistical model in ibid was not fitted directly against NWP model data, instead relying on a freely varying `regime fidelity' parameter to demonstrate that observed skill and signal-to-noise behaviour could be captured within a reasonable range of parameter values. We will show here, by explicitly fitting our statistical model, that the skill and signal-to-noise ratios of the European Center for Medium-range Weather Forecasts (ECMWF) seasonal prediction system (the Integrated Forecast System, abbr. IFS) can indeed be fully accounted for by its skill at capturing interannual variations in regime behaviour alone. This is done by basing our statistical model not on geopotential height regimes, but on the trimodal structure of the North Atlantic eddy-driven jet \citep{Woollings2010}. Because the NAO is to a large extent reflecting variations in the jet speed and latitude \citep{Woollings2010, Athanasiadis2010}, one may a priori expect to account for much of the model NAO skill in terms of its skill at predicting shifts in the jet speed and latitude. In fact, in \citet{Parker2019}, simple linear regression techniques were used to show that the majority of the IFS's skill at predicting the NAO could be attributed to its skill at predicting shifts in the jet latitude alone. However, this approach is only valid once the trimodality has been smoothed away by taking seasonal means, which renders the probability distribution Gaussian. Given the fundamentally trimodal structure of the jet latitude, we would argue that a more appropriate way to analyse the contribution from jet latitude variations is through regime dynamics. As explained above, in the framework of the regime model from \citet{Strommen2019}, the predictable component is not the average seasonal jet latitude itself, but rather the average seasonal persistence time-scales of the jet latitude regimes (which in turn determine the preferred jet latitude regime). By truncating the trimodal structure to a two-state system, we show that the IFS has skill at predicting seasonal shifts in these persistence time-scales, and hence that the signal-to-noise paradox can be explained by NWP models having systematic biases in regime persistence, as proposed in ibid. The success of such a two-state truncation is consistent with the hypothesis of \citet{Woollings2008}, which also viewed the NAO via a two-state system using the jet.

Secondly, having in this way fitted a regime dynamics model to jet latitude data for both reanalysis and the IFS, we analyse external sources of interannual variability of regime persistence. We show that commonly cited predictors of the NAO, such as ENSO, Kara sea-ice and the stratospheric polar vortex all project onto regime persistence. Of particular note here is the role of ENSO, which, while long thought to influence the NAO, does not have any significant linear correlation with the NAO index, implying that any such influence is necessarily non-linear. We demonstrate significant linear correlations of ENSO (via the Nino 3.4 index) with regime persistence time-scales, implying that regime dynamics may be a practical way to quantitatively measure the non-linear influence of ENSO on the NAO.

Finally, the link between NAO variability and persistence time-scales of the eddy-driven jet immediately suggests that a critical role may be played by local dynamics (as opposed to remotely driven forcing), in the form of transient baroclinic eddy feedbacks. There is a vast literature demonstrating the role played by such feedbacks in maintaining jet anomalies \citep{Hoskins1983, Shutts1983, Robinson1996, Lorenz2001, Lorenz2003, Zurita-Gotor2014}, and it is therefore reasonable to expect these to influence the persistence time-scales of the jet regimes used in our statistical model. By defining a suitable metric based on the eddy momentum flux convergence, we show that this is indeed the case, and that this is in fact the dominant source of jet latitude variability. For this metric we find that individual IFS ensemble members cannot generally reproduce the observed feedback strength, with a large ensemble mean required to boost the signal; a similar inadequacy for GloSea5 was observed in \citet{Scaife2019}. This suggests that the signal-to-noise paradox may be driven not just by weak teleconnections but also a weak eddy feedback in NWP models.

The structure of the paper is as follows. In section 2 we describe the data used, provide basic details of the IFS model and make some basic comments and definitions of key metrics; this includes the `ratio of predictable components' (RPC) metric \citep{Eade2014}, which is used as a measure of the signal-to-noise ratio. In section 3, in order to keep the paper fairly self-contained, we provide a basic description of the regime dynamics model constructed in \citet{Strommen2019}. In section 4, we discuss the relation between the jet and the NAO on daily and seasonal time-scales, including the influence of the trimodal jet latitude structure, for both reanalysis and the IFS. This analysis is used to motivate the explicit statistical model fitting carried out in the same section. In section 5, we show the results of this fitting, and verify that the high skill and low signal-to-noise ratio (i.e. high RPC value) of the IFS can be explained using our statistical model. In section 6 we analyse the influence of both external forcing and local eddy feedbacks on regime persistence time-scales, both for reanalysis and the IFS. Finally, in section 7 we make our concluding remarks.

\section{Data and Methodology}

\subsection{Data}
\label{sec:data}

The IFS model data considered comes from a seasonal hindcast experiment covering the 20th century \citep{Weisheimer2017}. A 51 ensemble member seasonal forecast is initialised in every 1st of November from 1900 to 2010 and allowed to run for 4 months, thereby producing a December-January-February (DJF) prediction for every year in this period. The model used is version CY41R1 of the IFS. Its spectral resolution is T255, corresponding to roughly 80km grid spacing near the equator, with 91 levels in the vertical. The model is run in atmosphere-only mode with forced sea-surface temperatures (SSTs). Because the hindcasts cover the early part of the century where observations are limited, the reanalysis product ERA20C \citep{Poli2013} is used to initialise the model atmosphere; this only assimilates observations of surface pressure. Both ocean and sea-ice boundary conditions come from the HadISST2.1.0.0 dataset \citep{Rayner2003}. Further details can be found in \citet{Weisheimer2017}.

While this paper will focus on the recent period 1980-2010, where more comprehensive reanalysis datasets assimilating satellite data are available, we will use ERA20C as our primary source of observations for consistency. Therefore all wind and geopotential height data used is from ERA20C, with SSTs and sea-ice coming from HadISST2. The exception for this is for the QBO, where the ERA-Interim dataset \citep{Dee2011} is used. This is due to the fact that ERA20C has a very weak QBO, owing to the fact that surface pressure data alone does not constrain the stratospheric circulation sufficiently. However, the QBO may have been partially forcing the surface pressure data (which \emph{does} severely constrain the NAO), so it is important to use a more realistic representation of it to assess its potential impact. 

The NAO is defined as the leading empirical orthogonal function (EOF) of detrended daily geopotential height anomalies at 500hPa (Z500), when restricted to the North-Atlantic region 90W-30E, 30N-90N. Anomalies are relative to a seasonal climatology computed across the full ERA20C time-period 1900-2010. The daily NAO of an individual IFS ensemble member is obtained by projecting its geopotential height field onto the reference EOF from ERA20C. A winter NAO timeseries is obtained by taking the DJF mean of the daily timeseries. It should be noted that the principal component timeseries produced in this way is not equivalent to that obtained by taking the leading EOF of DJF Z500 means. This latter approach only captures variability across seasons, while the use of daily data will also capture variability within seasons. However, the EOF's themselves are almost identical in their spatial pattern, with both capturing the familiar NAO dipole.

The speed and latitude of the eddy-driven jet are computed as in \citet{Parker2019}. Daily zonal means of zonal winds are restricted to the region 0-60W, 15N-75N and 850hPa, before being smoothed with a 9-day running mean. For each day of the DJF season, the speed of the jet is identified as the maximum wind speed attained in this domain and the latitude of the jet is the latitude at which this maximum is located. As noted in ibid, this procedure produces qualitatively similar results to more complex methods using winds at multiple levels or further filtering.

In Section \ref{sec:external_forcing} we consider various external predictors. For ENSO, we use the Ni{\~{n}}o 3.4 index, defined as the average across the region 5S-5N, 190E-240E. Since studies show that the response in the North Atlantic typically lags ENSO by around 1 month \citep{Jimenez-Esteve2018}, we consider the mean of this index over the November-December-January (NDJ) period to capture the influence on the North Atlantic DJF season. A QBO timeseries is defined by taking the DJF mean of zonal equatorial (10S-10N) winds at 50hPa, zonally averaged. The stratospheric polar vortex is measured using the November zonal mean winds at 50hPa averaged between 60N-80N. Kara sea-ice is measured as the November sea-ice anomaly, relative to the period 1978-2016, in the box 30E-75E, 67N-80N. The data is detrended to remove the linear trend resulting from declining sea-ice. Note that for these last two predictors, November means only were used to avoid confounding issues resulting from interactions and feedbacks between these and the NAO. Because of the slow time-scales of both ENSO and the QBO, as well as their spatial separation from the center of NAO action, this was not deemed to be necessary in these cases.

\subsection{Methodology}
\label{sec:methodology}

We are interested in three key metrics assessing model skill, all closely related. We define $Corr(EnsMean, Obs)$ to be the Pearson correlation coefficient between the ensemble mean of some forecast and observations over a given time-period. We will refer to this informally as the \emph{actual predictability}. On the other hand, we define $Corr(EnsMean, Mem)$ to be the average correlation between the ensemble mean and individual members of the ensemble. This is sometimes referred to as the `potential predictability'. However, as the signal-to-noise paradox has demonstrated clearly, this metric is not necessarily a meaningful upper bound on skill, so we follow the convention of \citet{Strommen2019} and refer to this instead as the \emph{model predictability} of the system, as it measures the model's ability to predict itself.

The third metric of interest is the `ratio of predictable components', $RPC$, defined in \cite{Eade2014}. One first defines the notion of the `predictable component' of a given system as the square root of the total fraction of variance that can be predicted in that system. When applied to observations we obtain $PC(Obs)$, and when applied to a model we obtain $PC(Mod)$. The $RPC$ is then by definition
\begin{equation}
RPC_{true} = PC(Obs)/PC(Mod) \nonumber
\end{equation}
In \cite{Eade2014} they provide an estimate of $RPC$ amenable to computation via the following inequality:
\begin{equation}
RPC_{true} = \frac{PC(Obs)}{PC(Mod)} \geq \frac{\sqrt{Corr(EnsMean, Obs)^2}}{\sqrt{\sigma_{sig}^2 /  \sigma_{tot}^2}} \nonumber 
\end{equation}
where $\sigma_{sig}^2$ is the ensemble mean variance and $\sigma_{tot}^2$ is the average variance of individual ensemble members. We define the RPC estimate to be the absolute value of this last term:
\begin{equation}
RPC_{est} = \frac{Corr(EnsMean, Obs)}{\sqrt{\sigma_{sig}^2 /  \sigma_{tot}^2}} . \nonumber
\end{equation}
Note the absolute value is necessary because we are taking a positive square root. As noted in \citet{Strommen2019}, if all ensemble members are assumed to share the same variance (which would in expectation hold given an infinite ensemble), then $RPC_{est}$ is precisely equal to the ratio of actual predictability divided by model predictability. Hence this metric will be 1 if all ensemble members and observations behave as random draws from the same underlying distribution. Further discussion of this metric can be found in ibid. While $RPC_{est}$ is an estimate of the true $RPC$, we will without comment refer to this estimate simply as the $RPC$.

In terms of significance testing, 95\% confidence intervals are, unless otherwise stated, generated by a Monte Carlo method of explicitly creating $>1000$ estimates using the statistical model in question. This is the case for all metrics computed with the regime dynamics model constructed in this paper. For correlations between various observed time-series (such as external predictors as in Section \ref{sec:external_forcing}), Monte Carlo samples were generated by a bootstrap resampling method. In practice, intervals found in this way were not appreciably different from those obtained simply using the standard error of correlations, so for simplicity we have used these.

Finally, we will frequently be examining statistical properties of probabilities, including relationships between different time-series of probabilities. To facilitate this, we typically work with \emph{log-odds} rather than probabilities when manipulating or generating probabilities. From an assumed distribution of probabilities $P$ (taking values between 0 and 1), a new, more Gaussian distribution $\hat{P}$ (taking values between $-\infty$ and $+\infty$) is obtained via the log-odds transformation 
\begin{equation}
\hat{P} = log(\frac{P}{1-P}). \nonumber
\end{equation}
Specific values $\hat{p}_s$ drawn from this latter distribution correspond to the probability $p_s$ from $P$ given by
\begin{equation}
p_s = \frac{1}{1+e^{-\hat{p}_s}}. \nonumber
\end{equation}
Because the distributions of probabilities we work with are log-normal or reverse log-normal, the log-odds distribution will be approximately Gaussian and so linear regression techniques can be used meaningfully. Note that a distribution $Y$ is called reverse log-normal if $1-Y$ is log-normal. While we often and without comment treat log-odds and probabilities interchangeably in discussion, the distinction is made clear when this is important.

\subsection{Non-stationarity of Predictability and Metrics for Modern Period}
\label{sec:nonstationarity}

It has been noted that the skill the IFS has at predicting the winter NAO is non-stationary over the twentieth century \citep{OReilly2017}. Because there appears to be variability not only in the role of external forcing but also local regime dynamics, it would be ideal to examine the framework presented in this paper across multiple periods of the twentieth century. However, because this would complicate the presentation of the paper considerably, we will focus on the recent period 1980-2010 on questions of predictability. Estimates of distributions or parameters that are independent of skill are computed using the full time-period 1900-2010, while those clearly dependent on skill are computed using the period 1980-2010.

In this modern period, using the definitions of Section \ref{sec:data}, the IFS ensemble mean winter NAO has a correlation with the observed winter NAO of $0.49$. The model predictability is $0.19$ and the $RPC_{est}$ over this period is $1.80$. These are the numbers that we wish to explain.

\section{The regime-based link between the jet and the NAO}
\label{sec:jet_nao_links}

\subsection{The distributions of jet latitude and jet speed}
\label{sec:jet_distributions}

It is well known that the NAO, while often depicted using sea-level pressure/Z500 anomalies, can be viewed as capturing variations in both the speed and orientation of the eddy-driven jet \citep{Thompson2003, Woollings2008} In this view, days in which the NAO is in its positive phase (NAO+ days) typically correspond to the jet being shifted northwards while days in the negative phase (NAO- days) correspond to a southward shift (e.g. Figure 1 of \citet{Woollings2010}). 

\begin{figure}[p]
	\centering
    \includegraphics[width=0.8\textwidth]{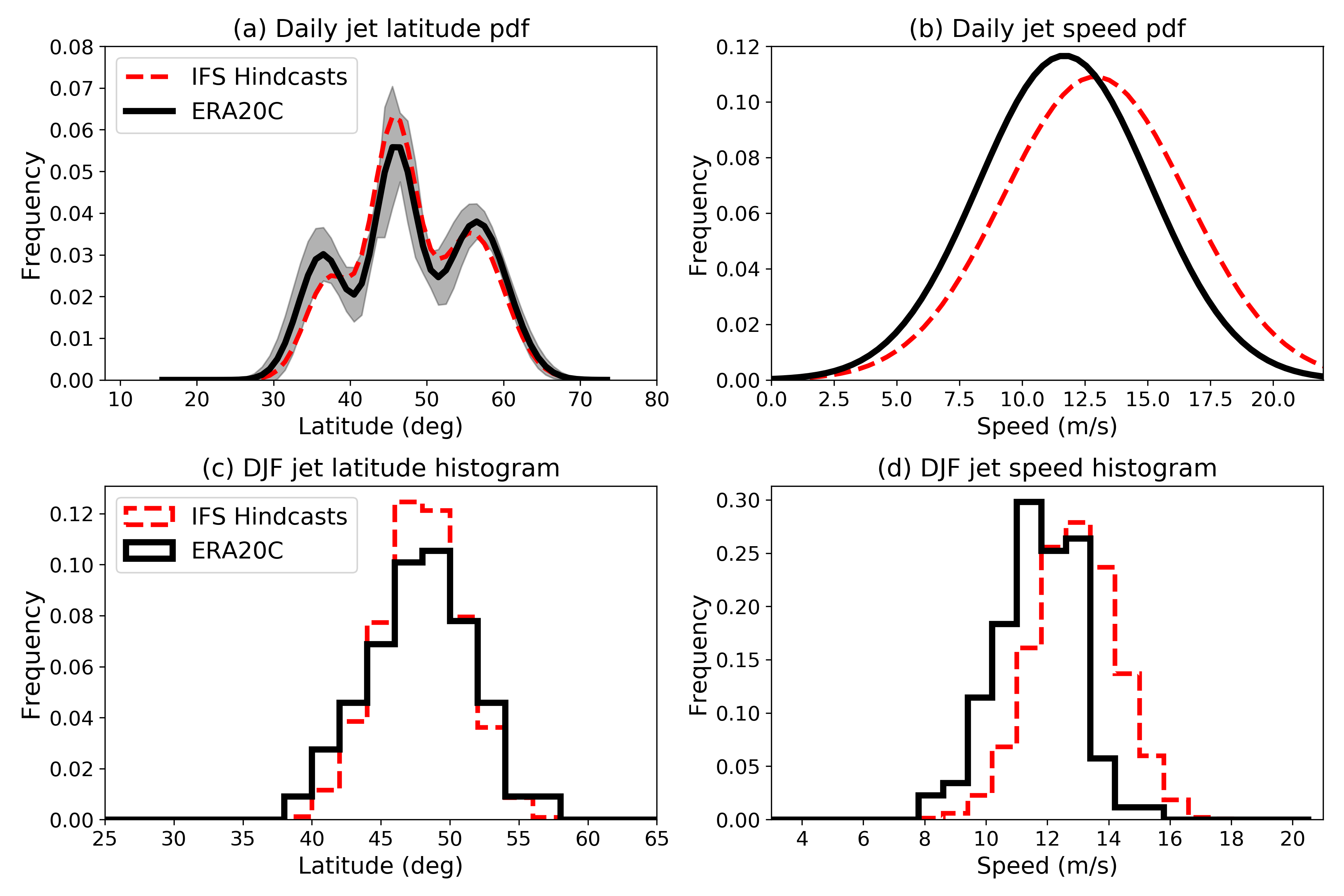}
	\caption{Estimated probability distributions of (a) the daily jet latitude; (b) the daily jet speed; (c) the DJF mean jet latitude; (d) the DJF mean jet speed. In all plots, ERA20C is shown in solid black and the IFS in dashed red.  The full period 1900-2010 is used.}
	\label{fig:jet_pdfs_both}
\end{figure}

By takings as its starting point an analysis of variations in the jet, as opposed to variations in the NAO, \citet{Woollings2010b} found that the jet tends to exhibit regime-like behaviour in terms of its location. Concretely, they identified three preferred locations, corresponding to the trimodality of the probability density function (pdf) describing the daily latitudinal position of the jet. A comparison of this pdf for ERA20C and the IFS was done in \citet{Parker2019}, which the reader should consult for a more comprehensive comparison. For completeness, we include a figure showing the pdfs here as well. To estimate a pdf from jet latitude data, we use a simple Gaussian mixture model assuming three peaks. As noted in ibid, the IFS also exhibits a trimodal jet, so this assumption on the mixture model is appropriate for the IFS as well. The estimated pdfs of both ERA20C and the IFS, using all 110 years of data, are plotted in Figure \ref{fig:jet_pdfs_both}(a). The three peaks are clearly visible and correspond closely to those in ibid. The shading indicates an estimate of uncertainty of the ERA20C probabilities, obtained by fitting the mixture model to individual 30-year chunks of the 110 years. While the locations of the peak are fairly stable, the relative sizes of each of the three modes vary more with time. Therefore, in line with ibid, we identify three distinct jet latitude regimes: Southern (15-40N), Central (40-52N) and Northern (52-70N). The IFS appears to be biased towards spending proportionally more time in the Central regime at the expense of the Southern and Northern regimes. By contrast, the pdf of daily jet speeds, shown in Figure \ref{fig:jet_pdfs_both}(b), is visibly Gaussian. The IFS has a clear bias towards stronger mean jet speeds than ERA20C.

\subsection{Daily and seasonal links between the jet and the NAO}
\label{sec:scatterplots}

In order to create a statistical model we need to establish a quantitative link between the jet and the NAO. Because we are ultimately interested in the DJF mean NAO, a first approach might be to relate the DJF means of the jet latitude and speed to the DJF mean NAO. This was the approach taken in \citet{Parker2019}, where the DJF NAO was decomposed into a jet speed and jet latitude component using multiple linear regression. Indeed, collapsing daily data to seasonal means completely eliminates the trimodal structure, rendering the pdf of seasonal mean jet latitudes normal: this can be seen in Figure \ref{fig:jet_pdfs_both}(c) and (d), showing histograms of DJF means of jet latitude and speed respectively. As multi-normality of variables is a requirement for linear regression, this smoothing away of the trimodal structure is necessary to make this approach work. By comparing these DJF decompositions of the NAO into a latitude and speed component for the IFS and ERA20C, ibid concluded that the IFS skill was, in this framework, attributable to the jet latitude skill alone. However, the fundamentally trimodal structure of the daily jet latitude suggests that this linear approach may not be the most appropriate and comes at the cost of removing structure which may contain valuable information.

\begin{figure}[p]
	\centering
    \includegraphics[width=0.8\textwidth]{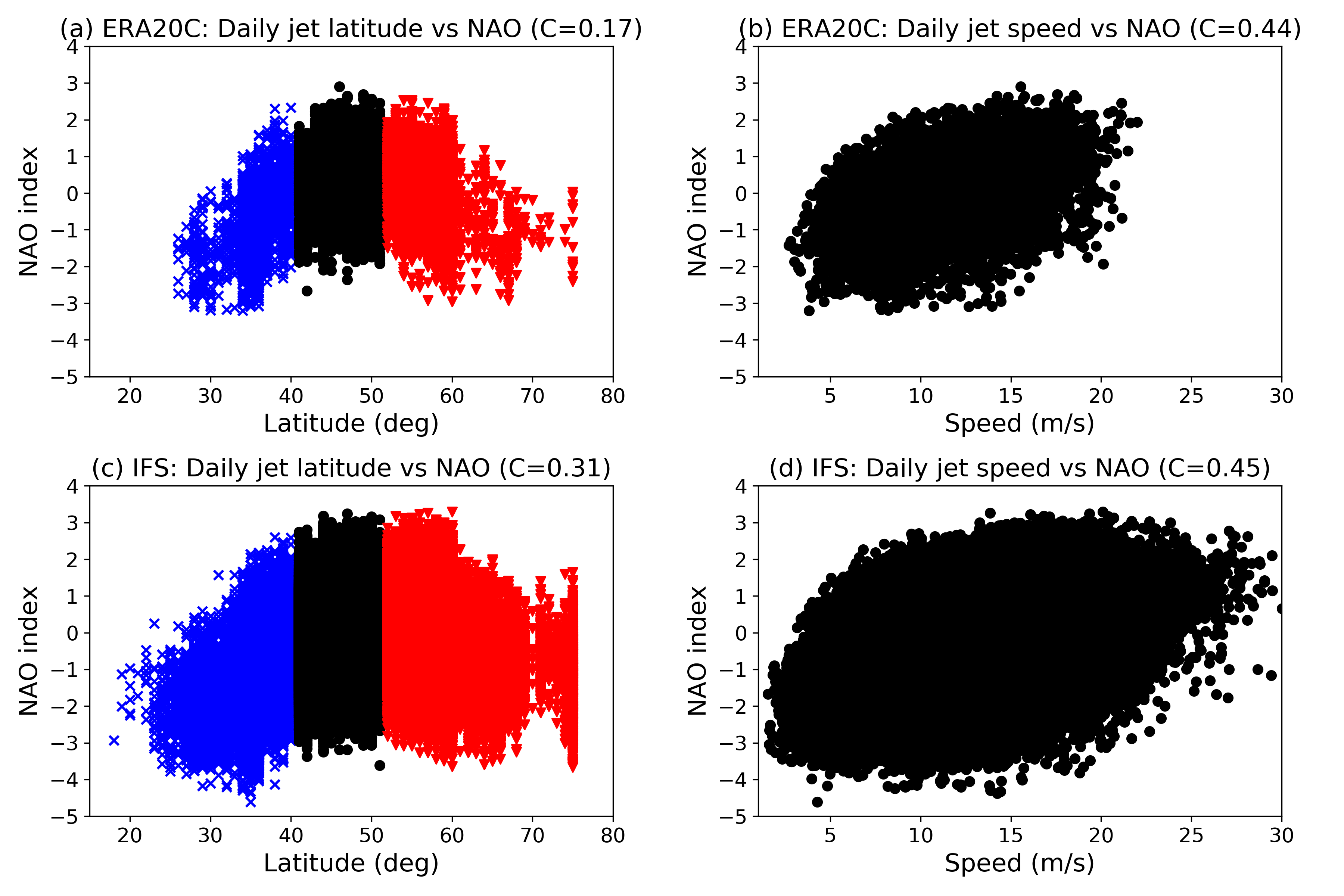}
	\caption{Scatter plots of daily jet quantities against daily NAO indices. In (a) and (c): ERA20C and IFS jet latitudes respectively. In (b) and (d): ERA20C and IFS jet speeds respectively. In each plot, the value $C$ in the title is the linear correlation between the two quantities. In (a) and (c), latitudes in the Southern jet regime (latitudes  $<$ 40N) are labelled with blue crosses; latitudes in the Central jet regime (40N $<$ latitudes $<$ 52N) are labelled with black dots; latitudes in the Northern regime (52N $<$ latitudes) are labelled with red triangles.}
	\label{fig:daily_jet_scatter}
\end{figure}

\begin{figure}[p]
	\centering
    \includegraphics[width=0.8\textwidth]{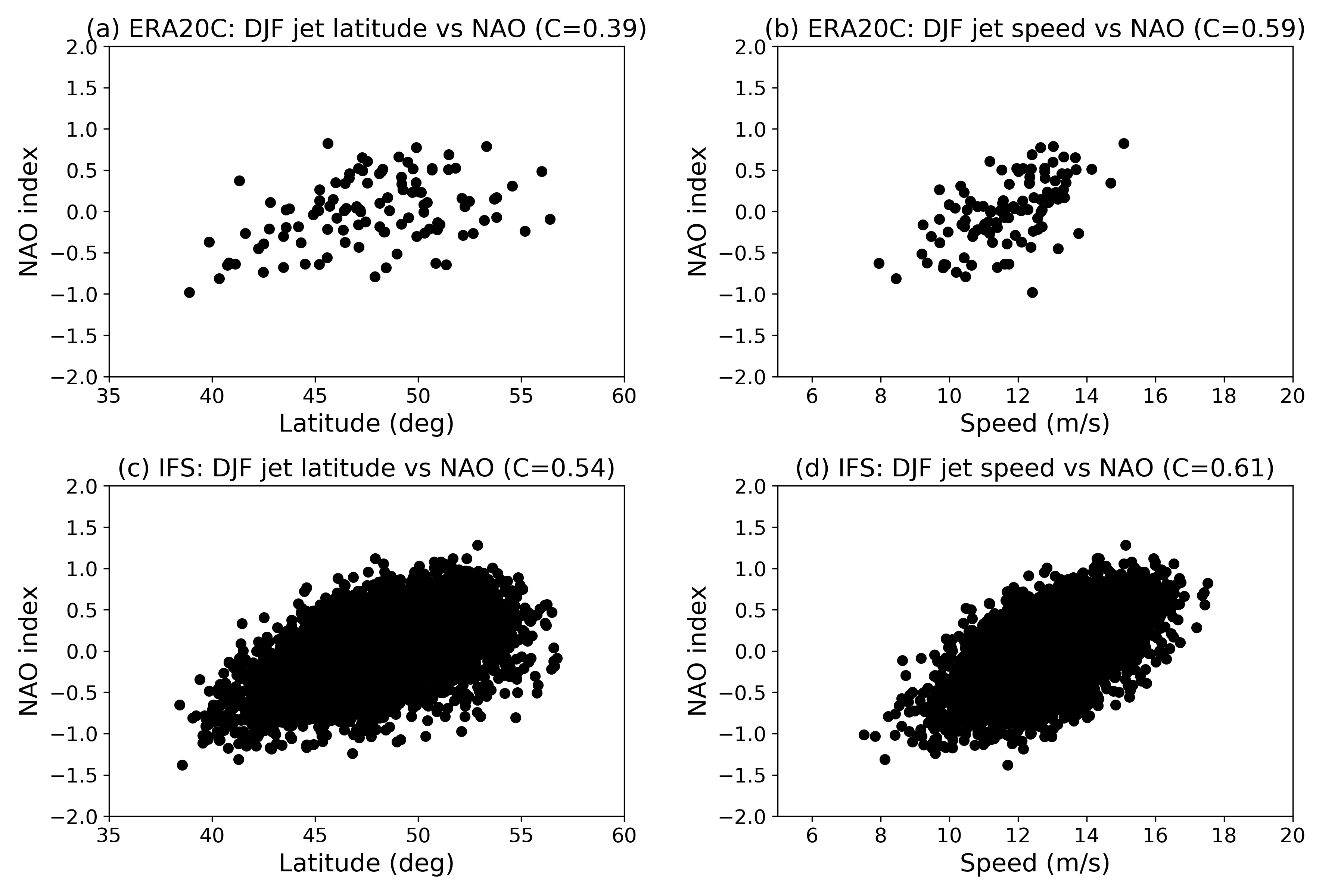}
	\caption{Scatter plots of DJF mean jet quantities against DJF mean NAO indices. In (a) and (c): ERA20C and IFS jet latitudes respectively. In (b) and (d): ERA20C and IFS jet speeds respectively. In each plot, the value $C$ in the title is the linear correlation between the two quantities. The full period 1900-2010 is used.}
	\label{fig:djf_jet_scatter}
\end{figure}

The failure of linear regression to account for the link between the jet and the NAO on daily time-scales can be seen clearly in Figure \ref{fig:daily_jet_scatter}, which shows scatter-plots of daily jet latitudes for ERA20C in (a) and the IFS in (c).  Similarly, daily jet speeds against daily NAO values are shown for ERA20C in (b) and the IFS in (d). While the relationship between the jet speed and the NAO appears to be linear in nature, the relationship between the jet latitude and the NAO is visibly non-linear, both for ERA20C and the IFS. In particular, negative NAO values may occur both at very low and very high latitudes. For convenience, the points in (a) and (c) have been marked according to which of the three regimes they are in. The three regimes each appear to have their own distinct relationship with the NAO; given the trimodality of jet latitudes, such a trimodal influence is perhaps to be expected a priori. By comparison, Figure \ref{fig:djf_jet_scatter} shows the same scatter-plots but using DJF means instead of daily values. While the relation between the jet latitude and the NAO now appears to be linear, the comparison with Figure \ref{fig:daily_jet_scatter} suggests that this is at least in part because the non-linear structure has been smeared away by averaging. Note that this non-linear relationship between the jet latitude and the NAO was also noted in \citet{Woollings2010}. Its potential implications for NAO predictability do not seem to have been commented on in the literature.

\begin{figure}[p]
	\centering
    \includegraphics[width=0.8\textwidth]{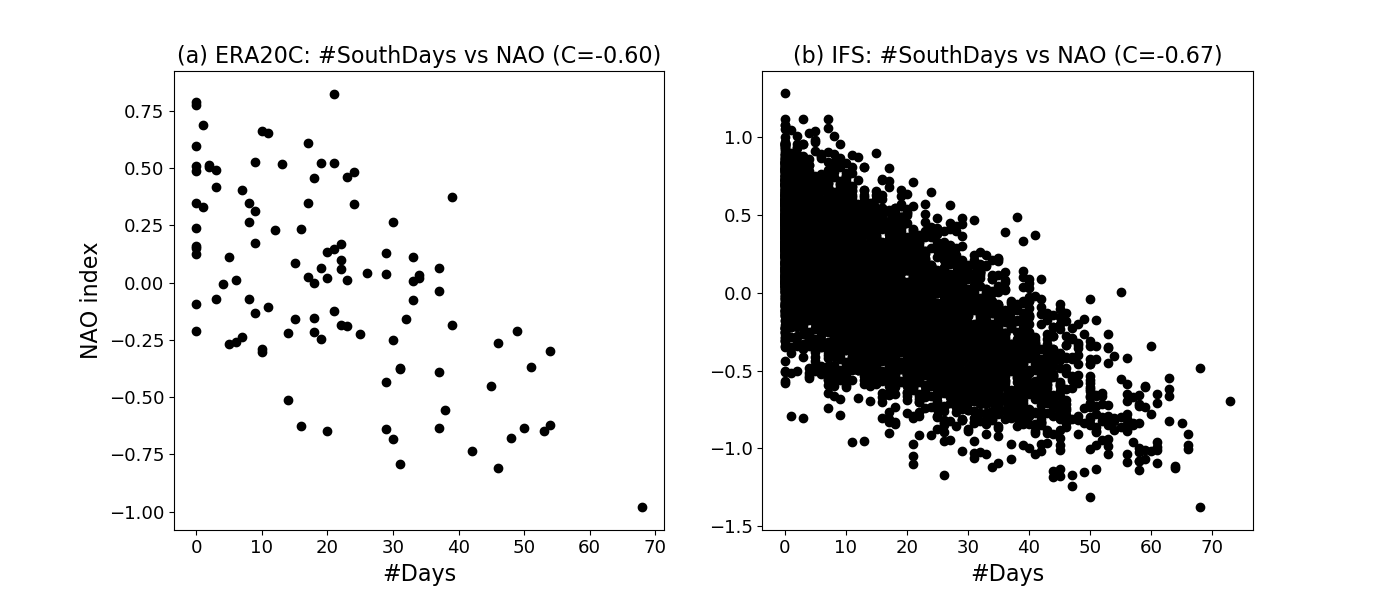}
	\caption{Scatter plots of the number of days per season spent in the Southern jet latitude regime against the DJF mean NAO. In (a) for ERA20C and (b) for the IFS. In each plot, the value $C$ in the title is the linear correlation between the two quantities. The full period 1900-2010 is used.}
	\label{fig:numdays_scatter}
\end{figure}

The non-linear relationship between the jet latitude and the NAO on daily time-scales is interpreted as evidence that a quantitative relationship between the jet and the NAO requires the framework of regime dynamics to capture fully. As noted in the Introduction, the predictable component across a given season is, in a regime system, the preferred regime state. A simple way to capture this is to simply count, in each DJF, the number of days spent in each regime. Figure \ref{fig:numdays_scatter} shows a scatter plot of the number of days spent in the Southern regime each season against the seasonal NAO mean, for ERA20C (a) and the IFS (b). There is a strong linear relationship, with a correlation of -0.6 for ERA20C and -0.67 for the IFS. The relationship is such that seasons where the jet spends a lot of time in the Southern regime correspond on average to NAO- seasons, consistent with the aforementioned finding in \citet{Woollings2010b} that NAO- seasons correspond on average to a southward shift in the jet. The correlations obtained for this regime-inspired metric are notably higher than those obtained by simply correlating DJF means of jet latitude ($-0.6$ vs $0.39$), implying that representing the regime-like behaviour of the jet can add quantitatively practical information. We will show in the next section that this relationship can be inferred to good approximation from the daily jet latitude structure.

Similar scatter-plots of the days spent in the Central/Northern regime against the winter NAO do not show as strong a signal, with correlations of 0.45 and 0.14 respectively. In terms of the jet latitude influence on the NAO, the dominant signature appears to be the proportion of time spent in the Southern regime versus the Central and Northern regimes. Hence, while the jet latitude itself is trimodal, when considered in terms of its influence on the NAO, a bimodal truncation, obtained by conceiving the Central and Northern regimes as one, appears to be appropriate. This is consistent with the analysis of \citet{Woollings2010b}, which effectively showed that a two-state mixture model approximation of the NAO selects precisely this bimodal jet latitude truncation (c.f. Figure 14 in ibid).

\section{The Regime Dynamics of the Trimodal Jet}
\label{sec:regime_dynamics}

In the previous section we argued that the influence of the jet latitude on the NAO should be understood via the regime-behaviour visibly taking place on daily time-scales. We showed that while the jet latitude is trimodal, its influence on the NAO is well captured by a bimodal truncation. In this section we will use this to construct a statistical model aimed at capturing the predictable signal coming from this bimodal regime structure contained in the winter NAO index.

Note that the statistical model we construct will only use information from the jet latitude, ignoring the jet speed entirely. While Figures \ref{fig:djf_jet_scatter}(b) and (d) show there is a strong linear relationship between the DJF mean jet speed and the DJF mean NAO, with correlations of around $0.6$, the IFS does not have any skill at predicting average jet speeds. This was already a conclusion of \citet{Parker2019}, but for our purposes it is sufficient to note that the correlation between the DJF jet speed of the IFS ensemble mean and the DJF jet speed of ERA20C is $\approx -0.02$. Therefore we would not expect jet speeds to be making a contribution to skill.

\subsection{The Markov model}
\label{sec:markov_model}

In this section we recall the main features of the statistical model of bimodal regime dynamics constructed in \cite{Strommen2019}: we refer to this model informally as the `Markov model'. Readers are asked to refer to ibid for a more complete discussion of this model.

We are going to imagine the jet as a bimodal system operating on daily time-scales. From this perspective, a DJF season corresponds to approximately 90 days, where, for each day, the jet resides in one of two states. The natural mathematical framework for generating 90 days of jet regime states is a two-state Markov chain. We will show in the next section that the autocorrelation of daily jet latitudes are well captured with an AR(1) process, implying that the Markov property does hold in this case and that it is not necessary to consider memory beyond one day.

\begin{figure}[p]
	\centering
    \includegraphics[width=0.8\textwidth]{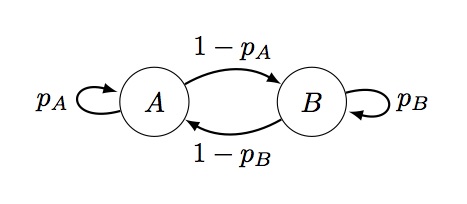}
	\caption{Illustration of a 2-state Markov chain with persistence probabilities $p_A$ and $p_B$ corresponding to states $A$ and $B$.}
	\label{fig:markovchain}
\end{figure}

Figure \ref{fig:markovchain} shows a schematic of a two-state Markov chain. The two states are currently denoted by the deliberately ambiguous monikers $A$ and $B$: they will be defined in the next section. The two states have persistence probabilities denoted by $p_A$ and $p_B$ (ranging between 0 and 1). If we imagine the chain operating on daily time-scales, this means concretely that if the jet is in the $A$ (resp. $B$) regime on a given day, it will remain in this regime the next day with probability $p_a$ (resp. $p_B$) or transition to the other regime with probability $1-p_A$ (resp. $1-p_B$).

To each state we now associate probability distributions which we denote by $X_{A}$ and $X_{B}$. These distributions represent the range of daily NAO index values that can be obtained when the jet is in states $A$ and $B$ respectively: they are assumed to be stationary. A single season of NAO values is now generated as follows. We assume that the probabilities $p_A$ and $p_B$ are fixed at the start of December and remain constant during the season. The jet will, on December 1st, be in one of the two states, after which it will carry out a 90 day random walk in the Markov chain. On each day when the jet is in state $A$, we draw a random number from the distribution $X_{A}$, which will be the NAO index value on that day. Similarly, on each day when the jet is in state $B$, the NAO index on that day is a random draw from $X_{B}$. At the end of the season, the 90 daily NAO values are then averaged to generate a DJF mean NAO.

If we let $\pi_{A}$ and $\pi_{B}$ denote the long-term average proportion of time spent in state $A$ and $B$ respectively, then one can show \citep{Strommen2019} that 
\begin{eqnarray}
\pi_{A} = (1-p_B) / (2 - p_A - p_B) \nonumber \\
\pi_{B} = (1-p_A) / (2 - p_A - p_B) \nonumber
\end{eqnarray}
and hence the preferred state is determined by the relative values of $p_A$ and $p_B$, with no preferred state if and only if $p_A=p_B$.  Furthermore, if $Y = Y(p_A, p_B)$ represents the distribution of DJF NAO means obtained through this process, it was shown in ibid that $Y$ has expected value
\begin{equation}
\mathbb{E}(Y) = \pi_{A}\mathbb{E}(X_{A}) + \pi_{B}\mathbb{E}(X_{B})
\end{equation}
and variance
\begin{equation}
Var(Y) = \frac{\pi_{A}Var(X_{A}) + \pi_{B}Var(X_{B})}{90}.
\end{equation}
Because the means $\mathbb{E}(X_A)$ and $\mathbb{E}(X_{B})$ of $X_A$ and $X_B$ are constant by assumption, this implies that the only source of predictability in a season comes from the persistence probabilities $p_A$ and $p_B$. These determine the values of $\pi_{A}$ and $\pi_{B}$ (i.e. the preferred state of the jet) which, via the above equations, determine the expected NAO value. Predictability of interannual variability in jet persistence therefore corresponds to predictability of interannual NAO variability. This predictability would be induced by seasonal deviations in the persistence probabilities from their climatological means, and the ability of NWP models to detect these. Possible triggers of such deviations will be considered in Section \ref{sec:drivers}.

A caveat to the above is that in seasons where $p_A$ and $p_B$ are close to each other, i.e. seasons where the jet only has a very weak preference to one particular state, then 90 days may not be a sufficient number of days to witness the above theoretical formulae, which become exact only in the limit of infinitely many days. Such a situation, in which the position of the jet may be dominated by internal variability, would correspond to a winter with a low level of predictability. Conversely, years in which $p_A$ and $p_B$ are strongly separated would correspond to a winter with a high level of predictability. We will see that, overall, there appears to be a sufficient amount of predictability in interannual variations of persistence probabilities to fully account for observed NAO predictability.

A final important point here is that, in principle, the probabilities $p_A$ and $p_B$ may be completely independent of each other. It is therefore necessary to consider the impact of any remote or local forcing on the system through both probabilities separately, with the overall impact being a combination of both. This will be analysed further in later sections.

\subsection{How predictable is jet latitude dynamics?}
\label{sec:jet_dynamics}

\begin{figure}[p]
	\centering
    \includegraphics[width=0.8\textwidth]{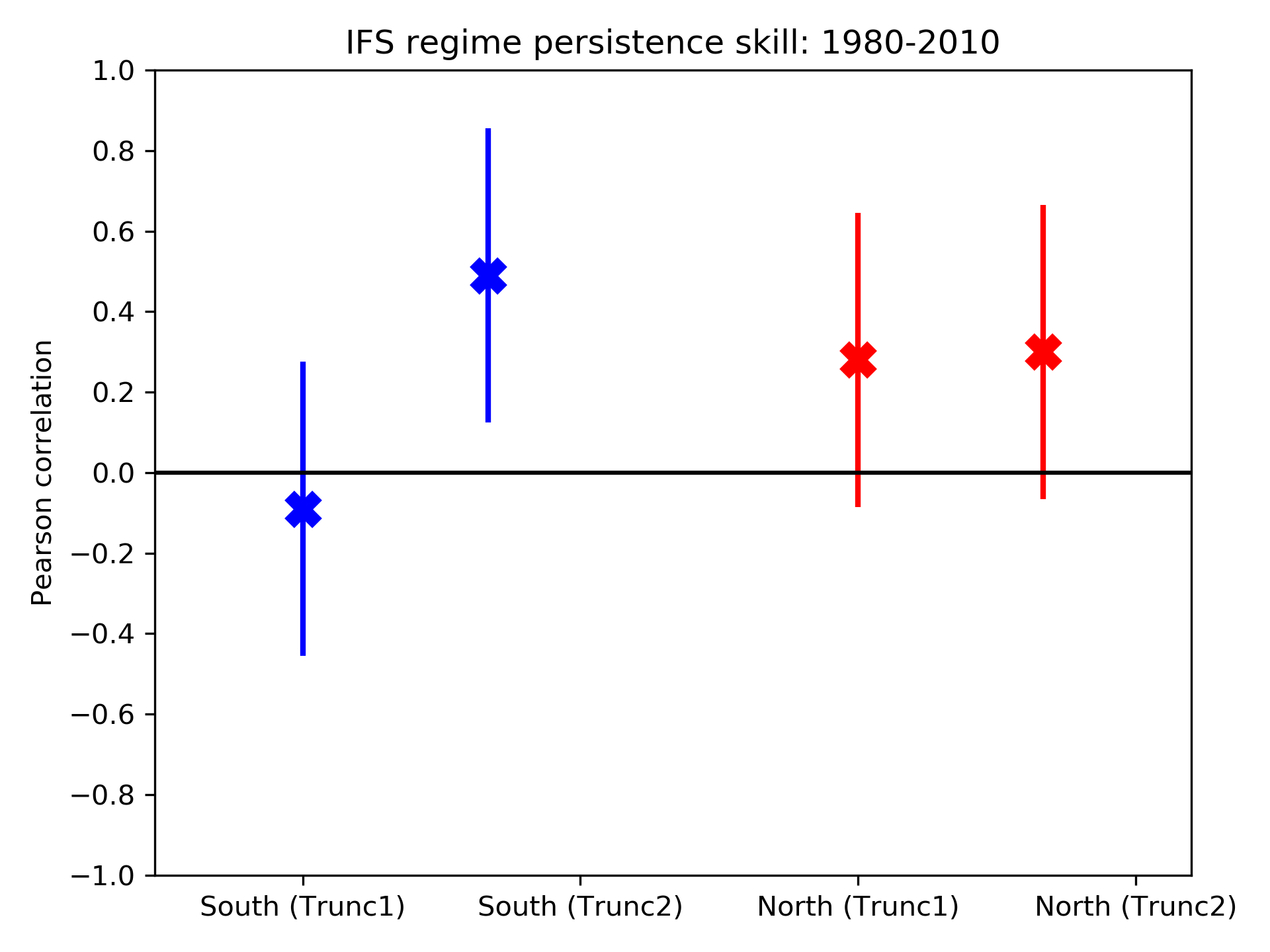}
	\caption{Correlations between the seasonal persistence probabilities of the IFS ensemble mean and ERA20C for the two regimes of the distinct bimodal truncations of the jet. `North/South (Trunc1)' correspond to a North regime obtained by concatenating the Central and Northern jet latitude regimes and a South regime corresponding to the Southern regime. `North/South (Trunc2)' correspond to a South regime obtained by concatenating the Southern and Central jet latitude regimes and a North regime corresponding to the Northern regime. Errorbars correspond to a standard correlation error of $\approx 0.36$. Data was restricted to the period 1980-2010.}
	\label{fig:correlation_skill}
\end{figure}

Fitting the above framework to actual jet latitude data requires first and foremost the identification of the two states $A$ and $B$. The conclusion of Section \ref{sec:scatterplots} suggests that we identify $A$ with the Southern regime and $B$ with the combined Central and Northern states. However, because we are ultimately interested in the question of predictability, we also need to take into account what the IFS does and does not have skill at predicting. To this end, we considered both potential bimodal truncations. Truncation 1 is defined as the one described above: a `South' regime corresponding to the Southern regime and a `North' regime corresponding to a combined Central and Northern regime. Truncation 2 is defined by having a `South' regime corresponding to the combined Southern and Central regimes and a `North' regime corresponding to the Northern regime. For each truncation, it is straightforward to estimate, for each regime, the seasonal persistence probabilities for both ERA20C and each IFS ensemble member. Figure \ref{fig:correlation_skill} shows estimates, for each regime, of the IFS skill at predicting the probabilities of ERA20C, measured as the correlation between the ensemble mean persistence and the persistence of ERA20C; only data from the period 1980-2010 was used. While the skill at predicting persistence of the North regime is largely insensitive to the truncation, with a correlation of $\approx 0.3$ either way, there is a dramatic difference for the South regime. For Truncation 1, the IFS has no notable skill at all, with a correlation of $\approx -0.1$. For Truncation 2 however, the IFS skill goes up to $\approx 0.5$. That is, while the truncation most relevant for determining the NAO is Truncation 1, the truncation actually predictable by the IFS appears to be Truncation 2.

While the Truncation 2 South correlation of $0.5$ is statistically significant, the corresponding North correlation of $0.3$ is not outside the $95\%$ confidence interval of the no-skill null hypothesis. To bolster confidence in both these correlations, we therefore considered the same correlations over earlier 30-year chunks. Concretely, we computed correlations over the thirty `chunks' of data 1950-1980, 1951-1981, ..., 1979-2009, 1980-2010. The mean across all these chunks are $\approx 0.55$ and $\approx 0.32$ for the South and North regimes respectively. In fact, these correlations are effectively stationary across each chunk, including for the very first period 1950-1980 which has no overlap with the modern period 1980-2010. The IFS therefore does appear to have robust levels of skill at predicting the persistence probabilities in Truncation 2, at least since 1950 onwards. Earlier periods begin to show evidence of the drop in NAO skill discussed in \cite{OReilly2017} and so were not included.

The fact that the IFS has skill only at predicting Truncation 2 is perhaps puzzling. However, there is evidence to suggest that the mechanisms driving jet variability are qualitatively different in the mid-latitudes, where the Southern and Central regime occur, compared to the higher latitude Northern regime closer to Greenland. In particular, \cite{Barnes2010} showed that transient eddy feedbacks acted to reinforce the jet in the mid-latitudes only, with northern anomalies rather capturing Rossby wave variability. If the IFS is simply able to broadly detect if jet variability is going to be dominated by baroclinic eddies or Rossby waves in a given season, then this may be expected to lead to skill at predicting persistence in the Truncation 2 context only, even if the IFS is not able to clearly distinguish between the Southern and Central regime. In Section \ref{sec:eddy_feedback}, we will independently confirm that eddy feedbacks appear to be crucial for understanding persistence of the Southern and Central regimes but less so for the Northern. Further analysis as to why IFS skill is limited to Truncation 2 will be left for future work.

Since it is Truncation 1 which is most relevant for determining the NAO, the lack of IFS skill for this truncation would at first sight appear to be a major obstruction towards explaining NAO skill using regime dynamics. However, distinguishing between a preference towards the Northern regime or the Southern and Central regimes can still be expected to contribute towards NAO skill, albeit to a lesser extent. We will see in Section \ref{sec:skill_estimates} that this contribution is never the less sufficient to account for the entirety of observed NAO skill.

\begin{figure}[p]
	\centering
    \includegraphics[width=0.8\textwidth]{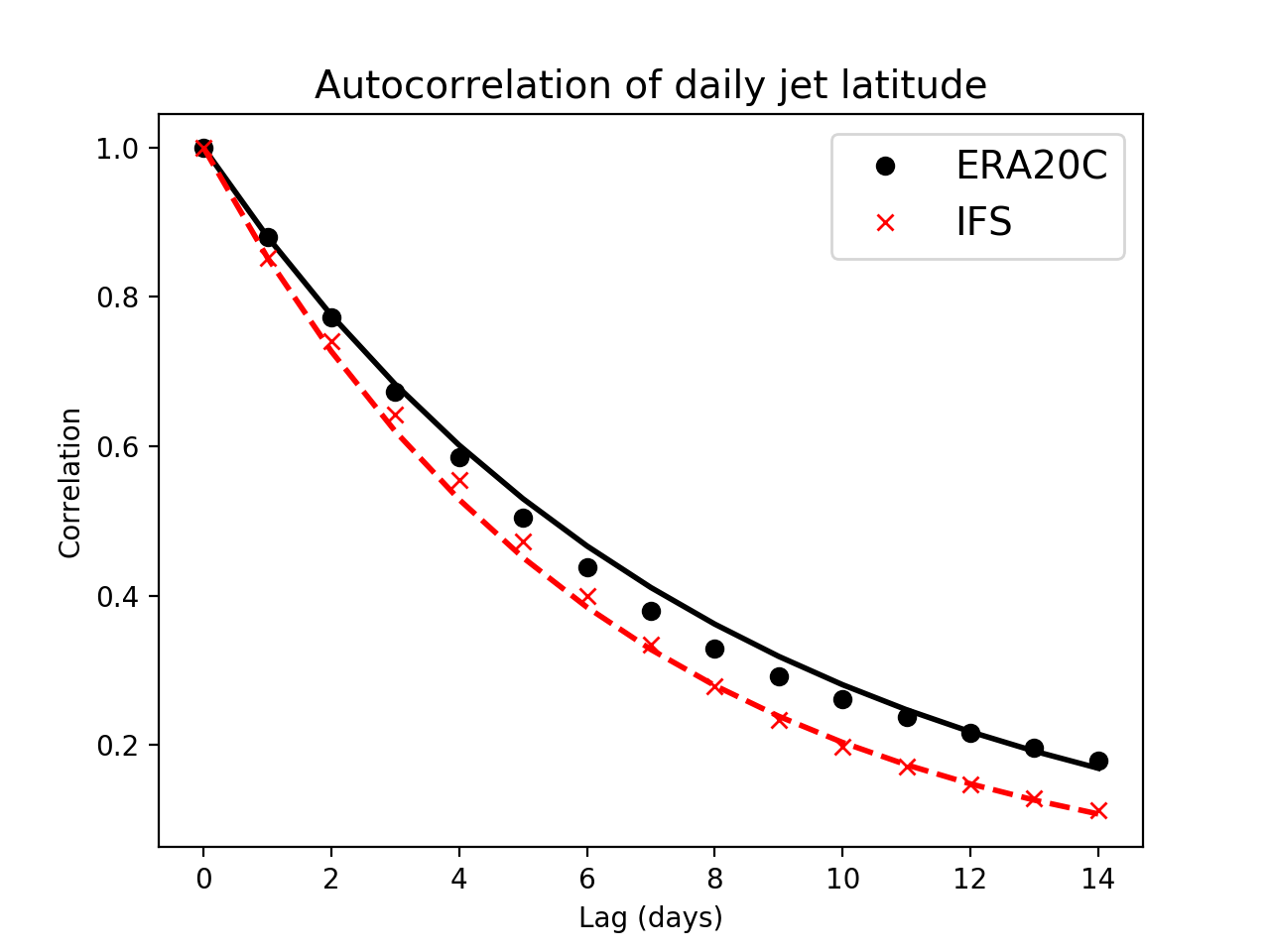}
	\caption{Autocorrelation of the daily jet latitudes of ERA20C (black dots) and the IFS (red crosses) for days 0 through 15. Solid lines show the theoretical autocorrelation functions for ERA20C (black solid) and the IFS (red stipled) under the assumption that the data is AR(1). The full period 1900-2010 is used.}
	\label{fig:autocorrelation}
\end{figure}

Finally, Figure \ref{fig:autocorrelation} shows the autocorrelation of daily jet latitudes for both ERA20C and the IFS, along with the theoretical autocorrelation functions assuming the data is AR(1) (i.e. satisfies the Markov property). The AR(1) functions match the observed data closely, confirming that the daily jet latitudes satisfy the Markov property to a sufficiently strong degree to justify the approach of Section \ref{sec:markov_model}. It is also clear that the IFS is in general under-persistent compared to ERA20C, with consistently reduced levels of autocorrelation.

\subsection{Determination of the Markov model distributions}
\label{sec:fitting_markov_model}

The analysis in the previous section suggests that we need to effectively be working with both Truncation 1 and Truncation 2 in the Markov model in a manner we will now sketch out. Truncation 2 (Southern+Central and Northern) will be used for determining the seasonal persistence probabilities of ERA20C and the extent to which the IFS can predict these. These will allow us to build up a quantitative measure of where the jet is spending most of its time each season. For each day spent (by ERA20C or the IFS) in the combined Southern+Central regime, we simply assume that the jet will be in the Southern regime according to the proportions implied from the full, trimodal jet latitude pdf: that is, with probability $\mathbb{P}(\text{Jet in Southern} \, | \, \text{Jet in Southern or Central})$. This sketch will be fleshed out fully in the next section, but we mention it now to better motivate the figures that follow.

\begin{figure}[p]
	\centering
    \includegraphics[width=0.8\textwidth]{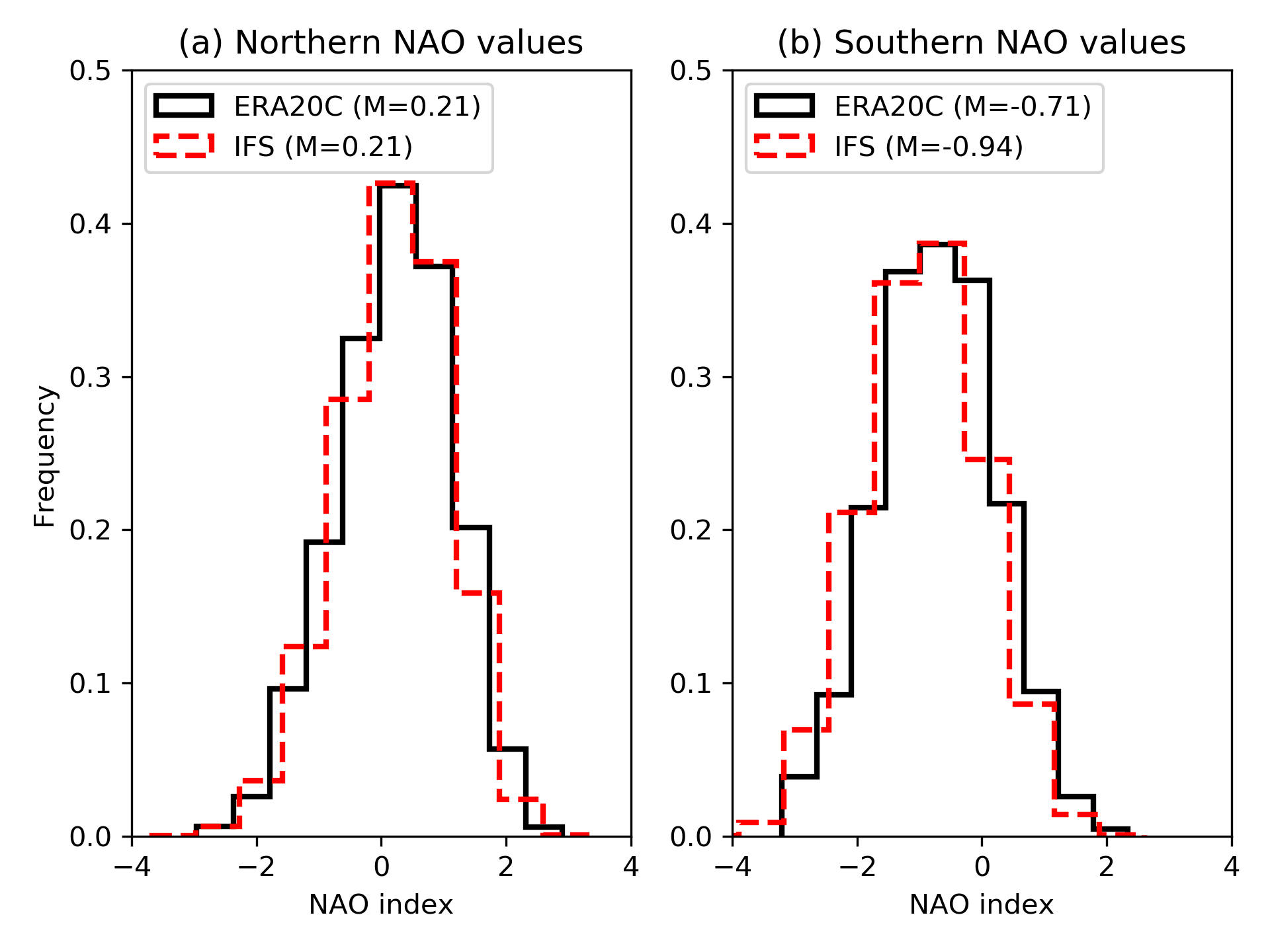}
	\caption{Estimated histograms of the possible NAO values obtained when the jet is (a) in the combined Central+Northern regime and (b) the Southern regime; ERA20C (black solid) and the IFS (red stipled). This corresponds to Truncation 1 in the main text. The values M in the legend denote the mean of the distributions. The full period 1900-2010 is used.}
	\label{fig:projections_trunc1}
\end{figure}

Figure \ref{fig:projections_trunc1} shows the estimates of $X_{A}$ and $X_{B}$ in Truncation 1 for both ERA20C and the IFS. Hence in (a) is shown the distribution of possible daily NAO values when the jet is in the Southern regime, and in (b) the same for when the jet is in either of the Central and Northern regimes. In concordance with earlier studies, the Southern jet is strongly associated with the NAO-, with ERA20C (resp. the IFS) having mean of $-0.7$ (resp. $-0.94$). The two combined regimes have a weaker association with the NAO+, with ERA20C and the IFS both having a mean of $0.21$. Hence, on average, regime dynamics would imply that if the jet spends $N$ days in the Southern regime, the DJF mean NAO for that winter would, for ERA20C and the IFS respectively, be
\begin{eqnarray}
\text{NAO}_{\text{ERA20C}}^{\text{Reg}} \approx \frac{-0.7\cdot N + (90-N)\cdot 0.21}{90} \approx -0.010\cdot N + 0.21 \\
\text{NAO}_{\text{IFS}}^{\text{Reg}} \approx \frac{-0.9\cdot N + (90-N)\cdot 0.21}{90} \approx -0.012\cdot N + 0.21.
\end{eqnarray}
On the other hand, directly fitting a linear relationship to the data in Figure \ref{fig:numdays_scatter} gives 
\begin{eqnarray}
\text{NAO}_{\text{ERA20C}}^{\text{Fit}} \approx -0.016\cdot N + 0.32. \\
\text{NAO}_{\text{IFS}}^{\text{Fit}} \approx -0.020\cdot N + 0.34.
\end{eqnarray}
Because the standard error of the slope coefficients in (5) and (6) is only around $0.002$, the mismatch between these two estimates seems robust for both ERA20C and the IFS. Indeed, it is unlikely that jet latitude dynamics alone can account for the full story of the jet and the NAO. One possible explanation for the mismatch is that in seasons where there is a strong forcing of the jet towards the Southern regime, this forcing may also be modulating the speed of the jet, which we know also influences the NAO independently of the latitude. Studies have also noted that the jet shows notable intra-seasonal structure which is not fully captured by the Markov model \citep{Franzke2011, Novak2015}. In any case, we will see that the agreement is close enough to allow regime dynamics alone to account for the skill of the IFS.

\begin{figure}[p]
	\centering
    \includegraphics[width=0.8\textwidth]{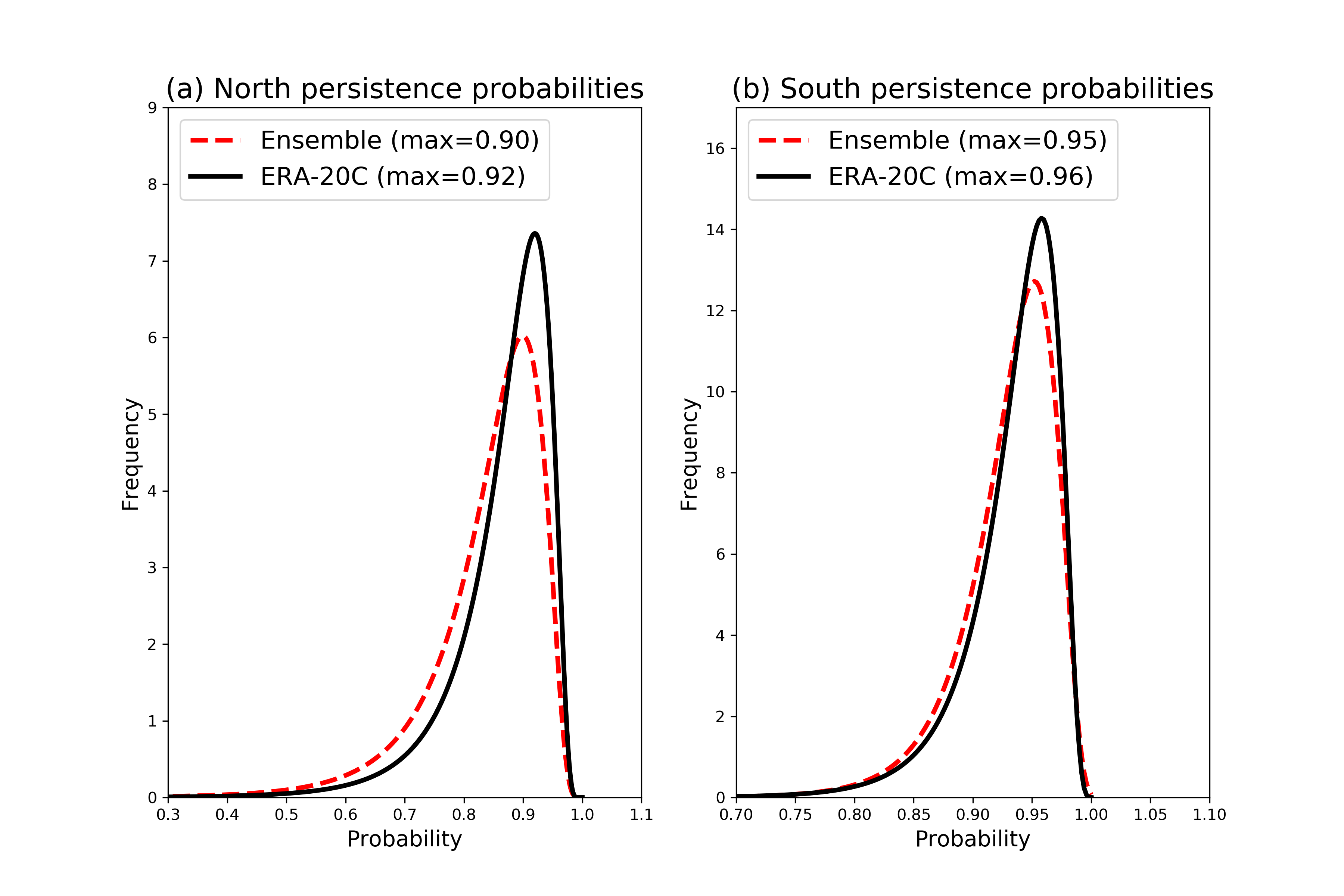}
	\caption{Estimated reverse log-normal distributions of seasonal persistence probabilities for (a) the combined Northern regime and (b) the combined Southern+Central regime; ERA20C (black solid) and the IFS (red stipled). This corresponds to Truncation 2 in the main text. The values of `max' in the legend denote the peaks of the distributions. The full period 1900-2010 is used.}
	\label{fig:persistence_trunc2}
\end{figure}

Figure \ref{fig:persistence_trunc2} shows estimates of the pdfs of seasonal persistence probabilities for Truncation 2. Hence in (a) is shown the distribution for the Northern regime, and in (b) is shown the same for the combined Southern and Central regimes. As in \citet{Strommen2019}, the histograms were evaluated by eye to be well captured by reverse log-normal pdfs. These were fitted to the data using the standard Python package scipy. It can be seen that both regimes show high levels of average persistence, with peaks close to or exceeding 90\%. The IFS shows a small but notable tendency towards under-persistence, particularly for the Northern regime. When similar distributions are fitted to each of the three jet latitude regimes individually (not shown) it is similarly found that the IFS is under-persistent for each. This is consistent with the reduced autocorrelation in Figure \ref{fig:autocorrelation}, suggesting that this tendency towards reduced persistence is a robust feature of the IFS. For further analysis of preferred transitions in the jet system, the reader may consult \citet{Franzke2011} and \citet{Novak2015}. Because our bimodal system is determined by the two persistence probabilities alone, such preferred transitions in the three-state system will not play a role in our paper.

The above information will allow us to generate seasonal persistence probabilities for ERA20C and link jet latitude regimes to the NAO. The final ingredient is to statistically represent the skill the IFS has at predicting the persistence probabilities of ERA20C, which we now address.

\subsection{Representing model skill and parameter fitting}
\label{sec:model_skill}

Our statistical model of NAO predictability will proceed as follows. Each winter season we draw two persistence probabilities for ERA20C, corresponding to the two regimes of Truncation 2, using the distributions shown in Figure \ref{fig:persistence_trunc2}. These are then used to generate a 90 day random walk across the system, which is finally translated, via the the distributions $X_A$ and $X_B$ of Truncation 1, into a DJF NAO mean. To simulate an ensemble hindcast of the IFS, we will also, each season, generate a pair of persistence probabilities for every ensemble member individually. In order to represent the observed IFS skill at predicting the persistence in ERA20C, the probabilities drawn for the ensemble members need to be correlated with the probabilities drawn for ERA20C. As a preliminary step towards this, we need to examine the extent to which the persistence probabilities of the two regimes are independent of each other.

\begin{figure}[p]
	\centering
    \includegraphics[width=0.8\textwidth]{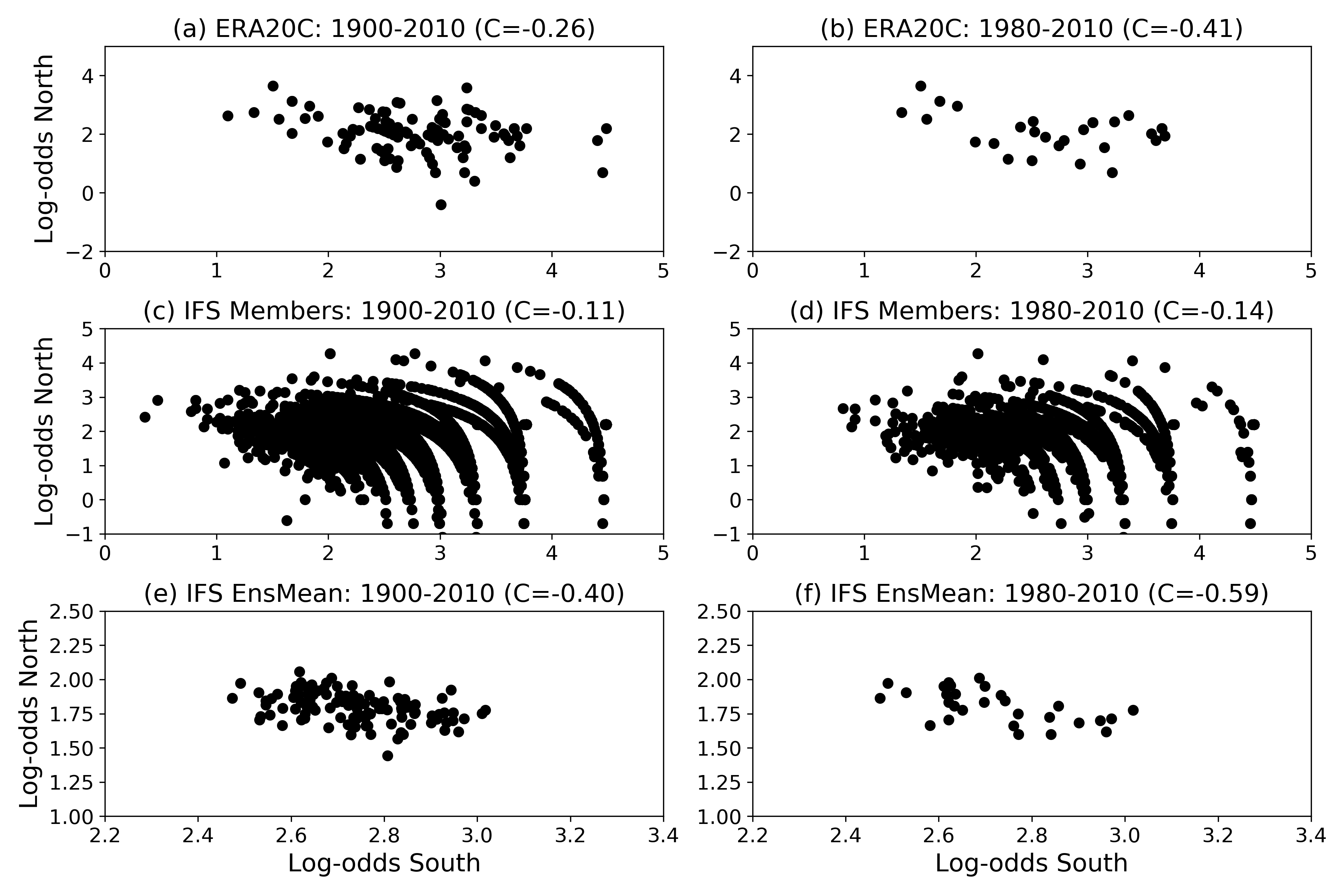}
	\caption{Scatter plots of the (log-odds of) persistence probabilities of the combined Southern+Central regime against those of the Northern regime. In (a) ERA20C (1900-2010), (b) ERA20C (1980-2010), (c) IFS ensemble members (1900-2010), (d) IFS ensemble members (1980-2010), (e) IFS ensemble mean (1900-2010) and (f) IFS ensemble mean (1980-2010). Note the ensemble mean persistence probability is the mean over all the ensemble member probabilities. The value $C$ in each plot's title is the linear correlation between the two quantities.}
	\label{fig:scatter_persprobs}
\end{figure}

Figure \ref{fig:scatter_persprobs} shows how the probabilities (shown as log-odds) of Truncation 2 (i.e. Southern+Central and Northern) are related. In (a), (c) and (e) are shown scatter plots of the Southern+Central probabilities versus the Northern probabilities for ERA20C, the IFS ensemble members and the IFS ensemble mean respectively. In (b), (d) and (f) are shown the same, but with data restricted to 1980-2010. The value $C$ in each denotes the correlation between the two quantities. The consistent negative sign of the correlation across all data suggests a link between the two probabilities. Part of this may be due to noise introduced from the fact that we always estimate probabilities over only 90 days, which may in some seasons be insufficient to obtain independent estimates of both. However, the fact that the correlation goes up significantly for the IFS ensemble mean, peaking at $\approx -0.6$ in the modern period, is taken as evidence that there is a robust signal here. Such a signal, if it existed, would naturally be amplified considerably by taking a large ensemble mean. As mentioned before, \citet{Barnes2010} showed that eddy feedback is not playing a major role for the Northern regime, suggesting that increased or reduced persistence there may simply be reflecting changes in mid-latitude persistence.

To formalise both the connection between the two probabilities and the IFS skill at predicting both, we utilise the framework of the `signal+noise' method from \citet{Siegert2016} as follows. Note that the following will all be phrased in terms of log-odds of probabilities.

Let $\hat{p}_{\text{obs}}$ be the persistence probability of regime $A$ for ERA20C in a given winter season and $\hat{p}_{\text{mod}}$ the corresponding probability for a single IFS ensemble member. Similarly, let $\hat{q}_{\text{obs}}$ and $\hat{q}_{\text{mod}}$ be the probabilities for regime $B$. Then we will assume that these four probabilities satisfy the following decompositions:
\begin{eqnarray}
\hat{p}_{\text{obs}} &=& s + \epsilon_{A} + \mu_{\text{obs},A} \\
\hat{p}_{\text{mod}} &=& \beta_{A} s + \eta_A + \mu_{\text{mod},A} \\
\hat{q}_{\text{obs}} &=& \lambda s + \epsilon_B + \mu_{\text{obs},B} \\
\hat{q}_{\text{mod}} &=& \beta_{B} \lambda s + \eta_B + \mu_{\text{mod},B}
\end{eqnarray}
Here the variables $s, \epsilon_{A}$ and $\eta_{A}$ are normally distributed with $s \sim \mathcal{N}(0, \sigma_{s}), \epsilon_A \sim \mathcal{N}(0, \sigma_{\epsilon_A}), \eta_A \sim \mathcal{N}(0, \sigma_{\eta_A})$, with the analogous notation for the $B$ variables. These are seasonally varying parameters, with $s$ representing the predictable component (i.e. signal) of $\hat{p}_{\text{obs}}$ and $\epsilon_A$ the unpredictable component (i.e. noise). We assume that $\hat{q}_{\text{obs}}$ is determined, modulo a noise term $\epsilon_B$, by $s$, via the scaling constant $\lambda$. This will ensure a degree of correlation between the two probabilities, the extent of which is determined by values of the parameters. The probabilities of the model are assumed to also be split into a signal and noise component. For regime $A$, the signal is assumed to equal a scaling factor $\beta_A$ times the true signal $s$. Hence $\beta_A$ is capturing the model's ability to capture the predictable signal. A value of $\beta_A < 1$ corresponds to the model damping the signal; $\beta_A = 0$ corresponds to no predictability, with $\beta_A=1$ corresponding to `perfect' skill at forecasting $\hat{p}_{\text{obs}}$. Similarly, the model captures the predictable signal $\lambda s$ of $\hat{q}_{\text{obs}}$ according to a scaling factor $\beta_B$. Finally, all the $\mu$ terms simply represent the climatological means of the respective distributions.

Conceptually therefore, we assume that in each DJF season there is precisely one predictable component in the jet system, namely a signal in how persistent the Southern+Central regime is going to be in reality. This completely determines, modulo unpredictable chaos, how persistent the Northern regime is going to be as well. The model is then able to capture the signal in both regimes according to fixed scaling parameters. The ability of the model to make good predictions thus depends on the relative size of the signal $s$ compared to $\epsilon$ (i.e. the signal-to-noise ratio of $\hat{p}_{\text{obs}}$), and the relative size of the scaled signal $\beta s$ compared to $\eta$ (i.e. the signal-to-noise ratio of $\hat{p}_{\text{mod}}$): similarly for $\hat{q}$.

It is clear that this is by no means the only way to represent the system, nor is any good fit of the model parameters unique. However, we will show that this representation, along with a very simple parameter fit, captures the main features of the system. As discussed in Section \ref{sec:nonstationarity}, we will fit parameters using the 1980-2010 data only, to ensure that we are not unfairly biasing our results. Parameter fitting is carried out using a `moments estimator' approach \citep{Siegert2016}, in which theoretical formulate of key statistical quantities (means and covariances) for the system defined by equations (7)-(10) are derived and compared to estimates from the data. The details are technical and can be found in the Appendix.

\begin{figure}[p]
	\centering
    \includegraphics[width=0.8\textwidth]{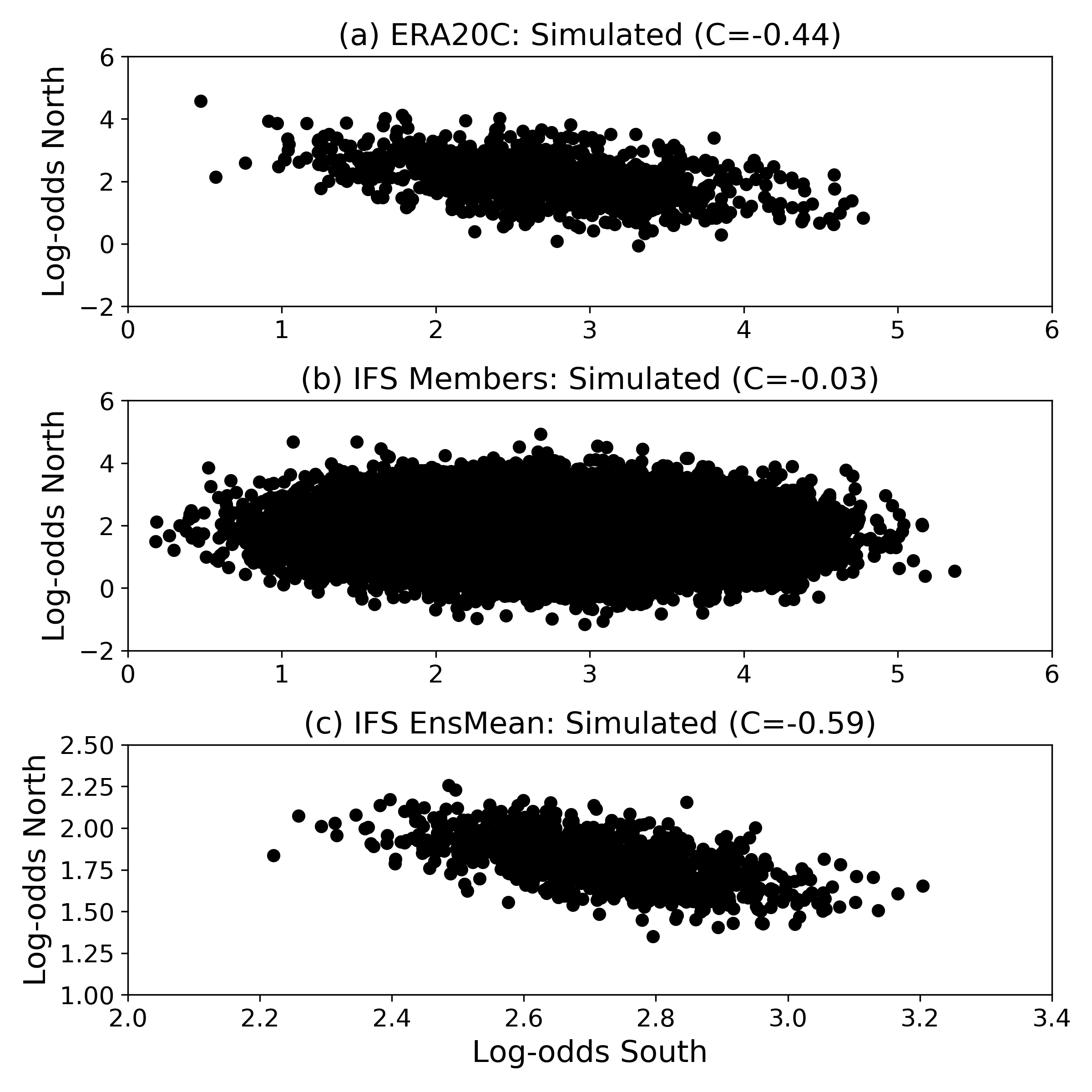}
	\caption{Scatter plots of sampled persistence probabilities, generated using equations (7)-(10) of the main text with the fitted parameters from Table \ref{tab:parameter_fits}. We generated 1000 samples of artificial ERA20C and IFS hindcast probabilities in this way, which are plotted in (a) for the simulated ERA20C, (b) the simulated IFS ensemble members and (c) the simulated IFS ensemble means. The value $C$ in the title of each is the linear correlation.}
	\label{fig:scatter_persprobs_samp}
\end{figure}

The parameters obtained are listed in Table \ref{tab:parameter_fits}. Using equations (7)-(10) with these parameters, we can generate a large sample of artificial probabilities. Such a random sample is shown in Figure \ref{fig:scatter_persprobs_samp} and should be compared to the right-hand column of Figure \ref{fig:scatter_persprobs}. It can be seen that the overall structure is recreated, including the means, variances and correlations, suggesting that the parameter fitting was effective. The parameter values themselves are physically sensible as well. That $\lambda \approx -1$ is saying precisely that any forcing which acts to increase the persistence of the Southern+Central regime will also result in an exactly corresponding decrease in the persistence of the Northern regime and vice versa. Both $\beta$ parameters are between $0$ and $1$, meaning that the model tends to dampen any forced increase in persistence, consistent with the conclusions of \citet{Strommen2019}. The noise terms of the IFS are slightly larger in amplitude compared to those of ERA20C, but this discrepancy is small compared to the dampened signal.

As a basic sensitivity test, we verified that the estimated parameter values were virtually unchanged when omitting any one of the 30 years between 1980 and 2010 before doing the fitting. We similarly found little change if fitting was carried out using the longer period 1970-2010. The inclusion of earlier periods began to show evidence of a decrease in skill (manifesting itself as smaller $\beta$ parameters) and so is not considered further.

Finally, by exploiting the fact that the mean of $\eta_{A}$ is zero, we can estimate the time-series of the persistence signal $s$ using the following relationship:
\begin{equation}
\overline{\hat{p}_{\text{mod}}} - \mu_{\text{mod},A}  \approx \beta_A s, \nonumber
\end{equation}
where overline denotes the ensemble mean. Dividing the quantity on the left-hand side by our estimated value of $\beta_{A}$ produces the desired signal, a key ingredient in our hindcast simulations using the Markov model.

\subsection{Simulating a seasonal hindcast}
\label{sec:simulation}

We now have all the ingredients necessary for simulating an IFS hindcast of the NAO during the period 1980-2010 using our Markov model. The extent of skill found in such a simulation will give an estimate of how much NAO skill can be accounted for purely through the regime dynamics of the daily jet latitude. The procedure for simulating a \emph{single winter season} proceeds as follows.

\begin{itemize}
\item[1.] Pick the estimated value of $s$ for the year in question. 
\item[2.] Draw random values of $\epsilon_A, \epsilon_B, \eta_A, \eta_b$.
\item[3.] Use equations (7)-(10) with all the above selected values to generate persistence probabilities for both ERA20C and the IFS.
\item[4.] Initialize both ERA20C and the IFS to start in either the Northern or combined Southern+Central regime. Let both of them perform a 90-day random walk for through the Markov chain specified by the persistence probabilities generated in step 2.
\item[5.] On days when either ERA20C or the IFS land in the Southern+Central regime, randomly specify it as having landed in the Southern or Central regime according to the relative proportions of the two regimes.
\item[6.] In this way, generate 90 days of residency in the full trimodal jet latitude structure.
\item[7.] Count the number of days spent in the Southern regime for both ERA20C and the IFS. Use equations (3) and (4) to generate a DJF NAO mean value for both.
\end{itemize}

\begin{figure}[p]
	\centering
    \includegraphics[width=0.8\textwidth]{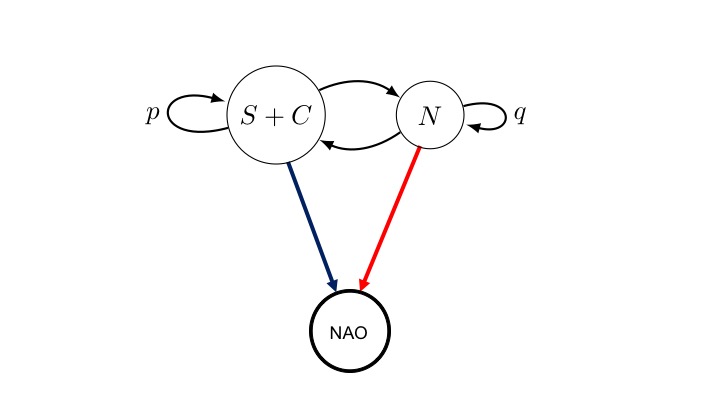}
	\caption{Schematic of the Markov model. The two jet-states are a combined Southern (S) and Central (C) regime and the usual Northern (N) regime. These have associated persistence probabilities $p$ and $q$ which vary annually in a manner which is partially predictable. Days when the atmosphere is in each state are translated into NAO index values (depicted as the blue and red arrows) using the distributions from Figure \ref{fig:projections_trunc1}.}
	\label{fig:schematic}
\end{figure}

\noindent Repeat this procedure 30 times to generate a full 30-year hindcast simulation of the period 1980-2010. For every such simulation, both ERA20C and the IFS ensemble members will have differing persistence probabilities, due to the random component in step 2, but all will have the same underlying signal estimated at the end of the previous section. A schematic of the Markov model is shown in Figure \ref{fig:schematic} to provide visual assistance to the reader.

In step 4, we note that the conditional probability of the jet being in the Southern regime given that we know it is in the Southern+Central combined regime is $\approx 33\%$ for both ERA20C and the IFS. Hence when either ERA20C or the IFS land in this combined regime, we specify it as having landed in the Southern regime with that probability. 

We would like to emphasise that while we have endeavoured to fully motivate our choices in the creation of this statistical model, the final results are fairly insensitive to many of the particulars. For example, equation (3) can be used to generate NAO values for both ERA20C and the IFS with no real change. Furthermore, the use of equations (3) and (4) themselves assume that the daily impact of the jet latitude on the NAO is always `average' in magnitude. If instead we draw \emph{random} daily values from the estimated distribution to generate more variability in these daily impacts the results are still broadly similar, because this variability is mostly averaged away when combining results from multiple hindcast simulations. We chose to use the average values to isolate the impact of jet latitude variations alone. As was the case in \citet{Strommen2019}, most sensible ways of implementing the Markov model are likely to produce qualitatively similar results.

\section{Estimating NAO skill with regime dynamics}
\label{sec:skill_estimates}

Using the methodology outlined in Section \ref{sec:simulation} we can now simulate a large number of hindcasts and estimate both the expected level of correlation and signal-to-noise ratios (i.e. RPC values). 
Concretely, we repeat the procedure outlined in the previous section 10000 times to produce this number of estimates of our desired metrics. These are then used to generate distributions of correlations and RPC values.

Carrying this out gives the following estimates:
\begin{eqnarray}
Corr(EnsMean, Obs) &=& 0.51 \pm 0.20 \nonumber \\
Corr(EnsMean, Mem) &=& 0.17 \pm 0.06 \nonumber \\
RPC_{est} &=& 2.10 \pm 0.91,  \nonumber
\end{eqnarray}
where the uncertainty estimates are two standard deviations around the mean.
Recalling the actual values from Section \ref{sec:nonstationarity} of 0.49, 0.19 and 1.80 , we see that there is very good qualitative agreement, with all three metrics well within the range of sampling variability. In particular, all the IFS skill at predicting the NAO can be accounted for purely through its ability to predict jet latitude persistence in a regime dynamics system.

\begin{figure}[p]
	\centering
    \includegraphics[width=0.8\textwidth]{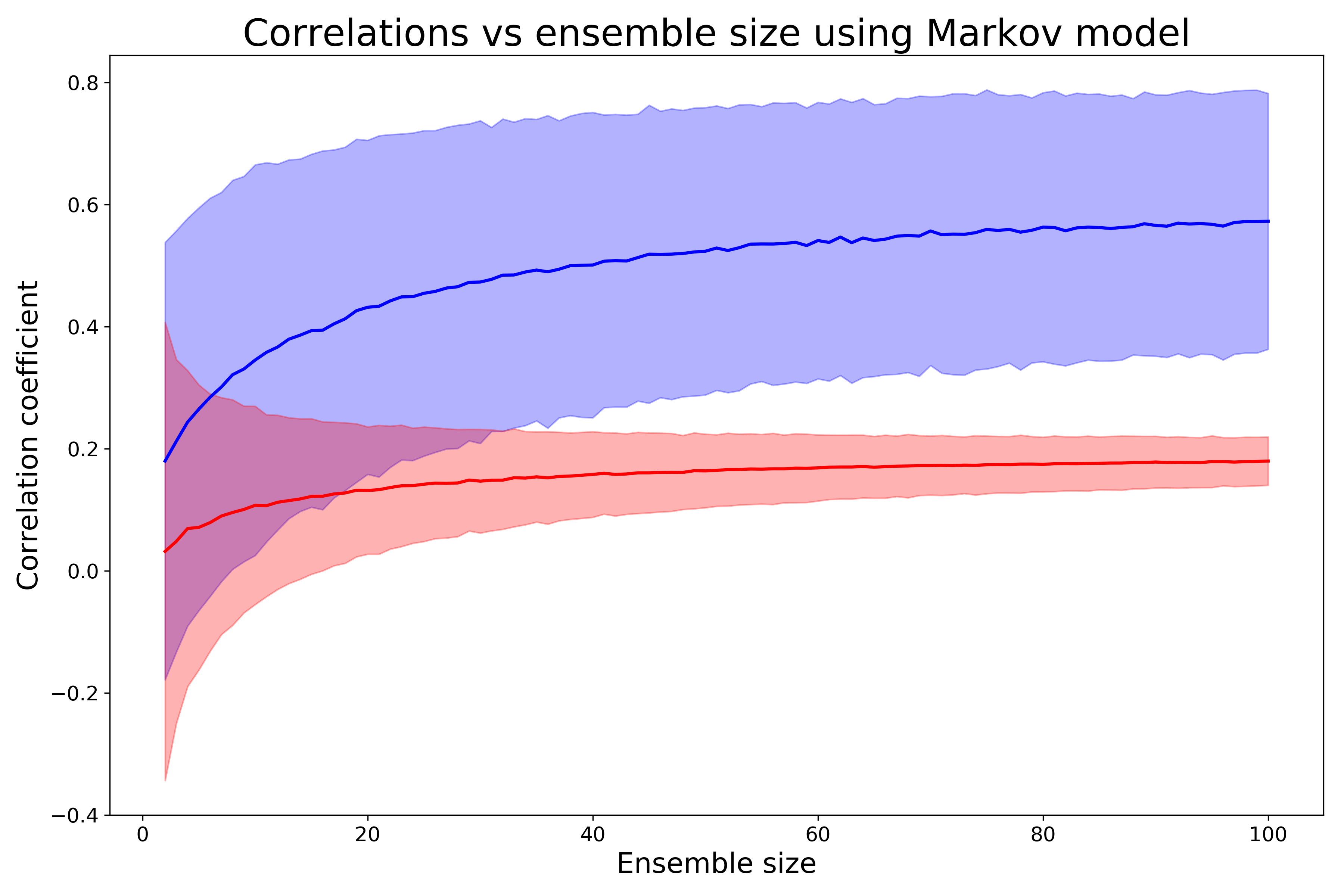}
	\caption{Estimates of expected actual predictability (upper, solid blue) and model predictability (lower, solid red) as a function of ensemble size, using the Markov model. Shading indicates two standard deviations from the mean. Each point uses 10000 simulations to estimate the mean and standard deviation.}
	\label{fig:skill_estimates}
\end{figure}

\begin{figure}[p]
	\centering
    \includegraphics[width=0.8\textwidth]{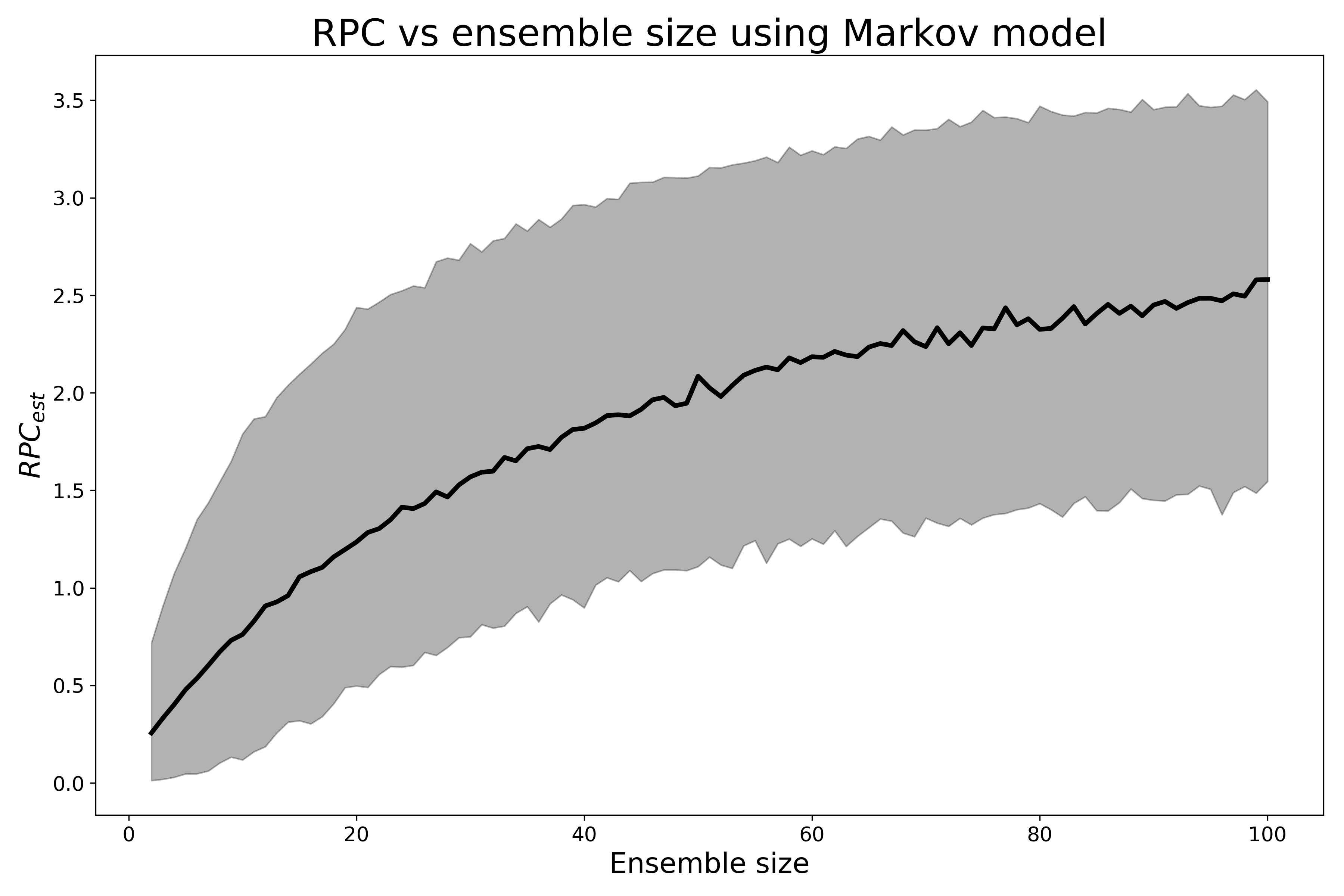}
	\caption{Estimates of RPC values using the Markov model. Shading indicates two standard deviations from the mean. Each point uses 10000 simulations to estimate the mean and standard deviation.}
	\label{fig:rpc_estimates}
\end{figure}

Figure \ref{fig:skill_estimates} shows how the actual predictability and model predictability vary as a function of the ensemble size. It can be seen that the actual predictability begins to asymptote towards $\approx 0.6$. If we set the $\beta$ parameters to 1, the expected ensemble mean correlation with 51 ensemble members is $\approx 0.62 \pm 0.18$; increasing the ensemble size to $1000$ does not lead to any increase, implying that this is the maximum expected correlation that can be achieved purely from jet latitude dynamics. Consequently, the proportion of NAO variance that can be accounted for by jet latitude persistence alone is around $0.62^2 \approx 0.38$. Figure \ref{fig:rpc_estimates} similarly shows how $RPC$ scales with ensemble size. This metric is more sensitive to ensemble size and does not begin to approach the theoretical limit of $0.51/0.16 \approx 3.2$ until the ensemble size approaches $1000$. Increasing the $\beta$ parameters to $1$ and using 1000 ensemble members brings the $RPC$ down to $\approx 1.05$, with the actual predictability at $0.62$ (as mentioned) and the model predictability at $0.60$. This implies that the low signal-to-noise ratio here is not due to the IFS having noisier regime dynamics ($\sigma_{\eta} > \sigma_{\epsilon}$) but is almost entirely accounted for by its damping of persistence ($\beta < 1$).

\begin{figure}[p]
	\centering
    \includegraphics[width=0.8\textwidth]{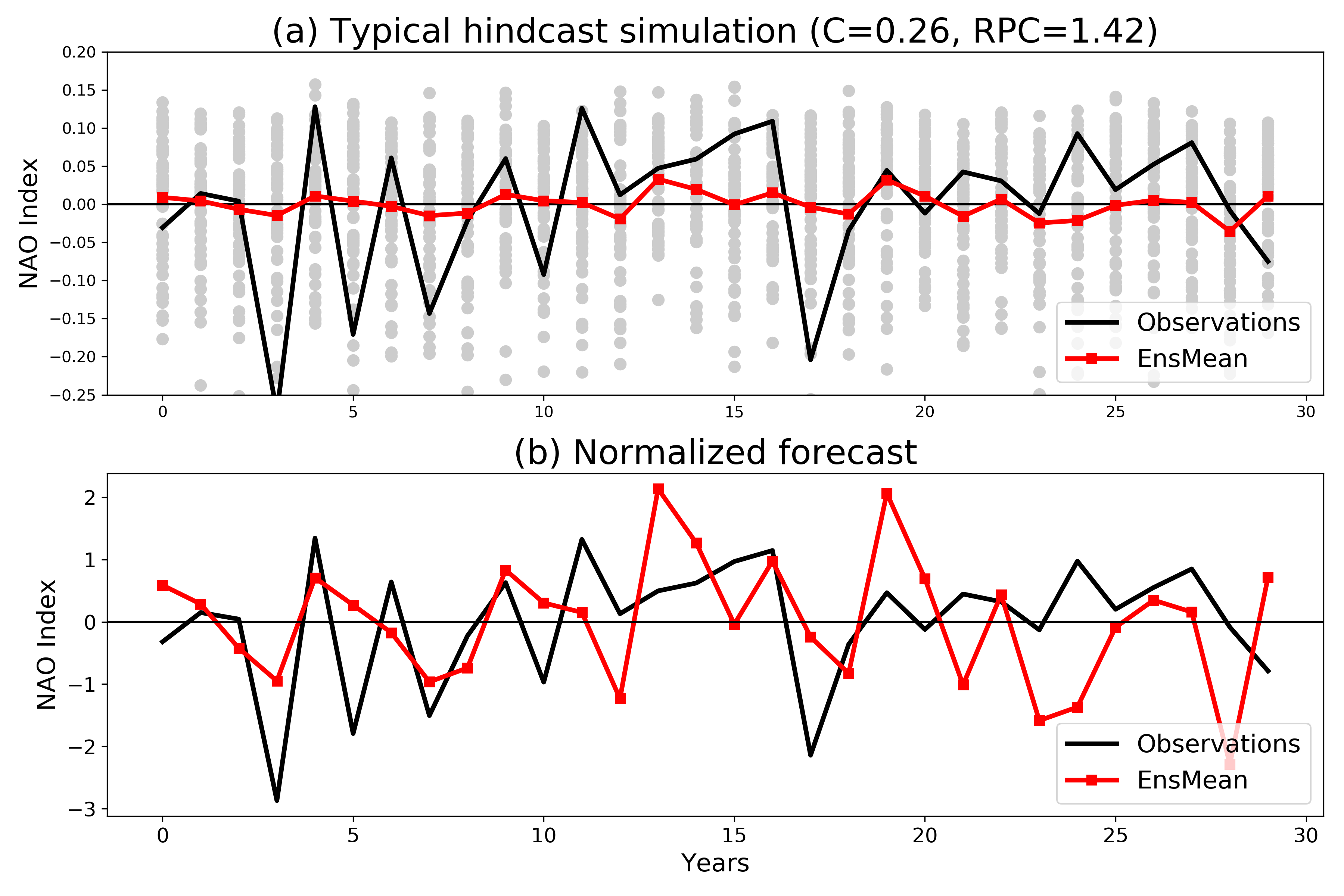}
	\caption{Example of a simulated NAO hindcast using the Markov model. In (a), simulated ERA20C (black solid), simulated IFS ensemble mean (red solid-dotted), with grey dots representing individual IFS ensemble members; in (b), the same simulated ERA20C (black) and IFS ensemble mean (red dotted) but normalized to have standard deviation 1. The value $C$ in the title of (a) denotes the ensemble mean correlation with the simulated ERA20C, with $RPC$ denoting the estimated $RPC$.}
	\label{fig:example_simulation}
\end{figure}

Figure \ref{fig:example_simulation} shows a typical hindcast simulation generated with the Markov model. In (a) is shown the raw hindcast, with the simulated ERA20C (black), IFS ensemble mean (red) and grey dots representing individual ensemble members. In (b) is shown the same simulated ERA20C and IFS ensemble mean but now with both timeseries normalized to have standard deviation 1. This should be compared to Figure 1 and 2 of \citet{Dunstone2016} or Figures 8 and 9 in \citet{Strommen2019}. The ensemble mean correlation and RPC in this case were $0.48$ and $2.07$ respectively. The ensemble variance is almost identical to that of ERA20C, which is a general feature of the Markov model. Note also the visible presence of negative skew in the ensemble. In fact,  this negative skew is a natural feature of the Markov model, owing to the higher climatological mean persistence of the Southern+Central regime compared to the Northern regime. It is well known that the winter NAO is significantly negatively skewed, implying that this skew can be explained naturally through regime dynamics.

While the correlations and RPC values, including their scaling with ensemble size, can also be reproduced with more linear techniques, as in \citet{Siegert2016} for the UK Met Office model, the Markov model has some key additional achievements. Firstly, it fully respects the trimodal structure of the jet, without relying on collapsing this structure by taking DJF means. Secondly, it straightforwardly produces a negatively skewed NAO, which (to our knowledge) cannot be achieved through linear techniques alone without artificially inserting the skew. Finally, by highlighting the role of daily time-scale jet persistence, the Markov model points towards a concrete physical mechanism driving predictability, namely transient baroclinic eddy feedback. This is discussed further in the next section.

\section{External and Internal Drivers of Regime Variability}
\label{sec:drivers}

Having demonstrated that the IFS skill at predicting the winter NAO can be attributed to its ability at predicting interannual variations in the extent of jet persistence, we now examine potential drivers of such interannual variability. First we examine external sources of forcing (teleconnections), before examining local mechanisms in the form of eddy feedback.

Because a comprehensive and in-depth study of predictors of regime dynamics would likely make the present paper excessively long, we keep the discussion brief. We aim to examine this more in future work.

\subsection{External Forcing}
\label{sec:external_forcing}

As discussed in the introduction, some of the most commonly proposed sources of external NAO forcing are sea-ice in the Kara sea region, ENSO (Nino 3.4) and the stratosphere, both in the form of the stratospheric polar vortex and the QBO. We will examine the influence of precisely these factors on seasonal persistence probabilities in both the Southern+Central and Northern regimes. We remind the reader that details of how these predictors are defined was included in Section \ref{sec:data}.

Note that while in our implementation of the Markov model, the entirety of the predictable signal comes from the Southern+Central persistence, this may well be an oversimplification. As mentioned already in Section \ref{sec:jet_dynamics}, it was shown in \citet{Barnes2010} that while persistence driven by eddy feedback was a key mechanism in the mid latitudes, variability further north is driven more by Rossby waves. This implies that the influence of external forcing may be different, and hence partially independent, for the two regimes.

\begin{figure}[p]
	\centering
    \includegraphics[width=0.8\textwidth]{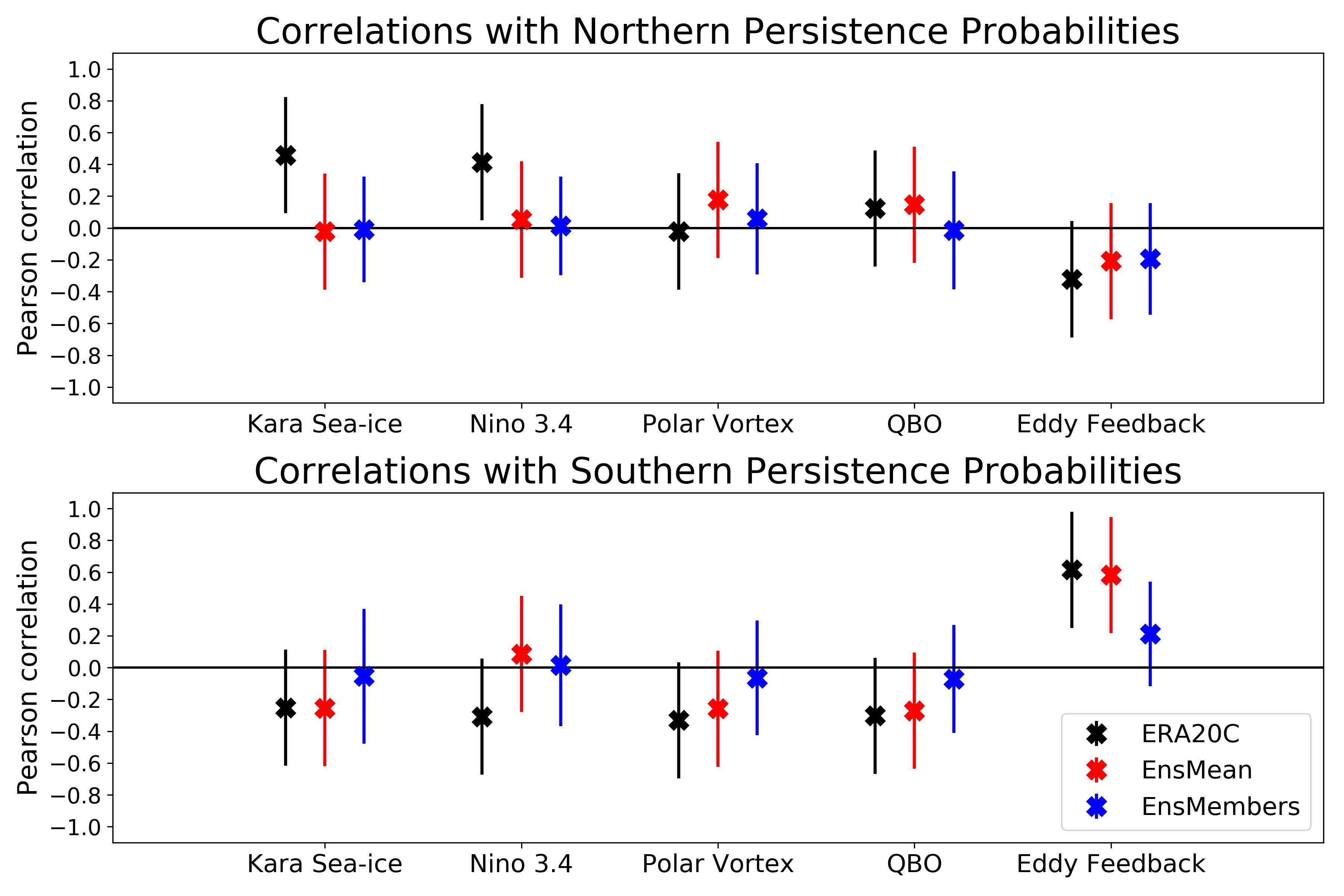}
	\caption{Correlations between various predictors and Northern persistence probabilities (top panel); Southern+Central persistence probabilities (bottom panel). Data is restricted to the period 1980-2010. For each predictor (labelled on the x-axis), the correlations are, from left to right, for ERA20C (black), the IFS ensemble mean (red) and the average across individual IFS ensemble members. The cross denotes the mean, with the spread representing twice the standard error for correlations of time-series of length 30, i.e $\approx 0.36$.}
	\label{fig:predictors_correlations}
\end{figure}

Figure \ref{fig:predictors_correlations} shows correlations between the various predictors and persistence probabilities of ERA20C (leftmost and black), the IFS ensemble mean (middle and red) and the average across individual IFS ensemble members (blue and rightmost); for the Northern regime (top panel) and the Southern+Central regime (bottom panel). In the North, the significant predictors appear to be Kara sea ice and ENSO. The overall largest influence in ERA20C comes from Kara sea ice, with a correlation of around $0.45$, in agreement with previous studies \citep{Wang2017}. More striking is the significant linear correlation, of around $0.4$, between ENSO and the Northern persistence probabilities here, which is far from obvious. Indeed, a major obstacle towards a \emph{quantitative} estimate of the influence of ENSO on the NAO is that, while studies suggest there is a relationship between the two, there is virtually zero linear correlation between the Nino 3.4 timeseries and the NAO index (see e.g. ibid). The significant correlation here suggests that jet latitude regime dynamics may offer a route towards such quantitative estimates of what proportion of NAO variance is explained by ENSO. However, while ENSO is influencing the persistence of ERA20C, the IFS does not appear to capture this at all, with neither the ensemble mean nor individual members showing any significant correlation. We remind the reader that because the hindcasts considered used forced SSTs, the Nino 3.4 index is identical for both ERA20C and all IFS ensemble members. Therefore this seems to be a failure in the propagation of the ENSO signal sufficiently far downstream. The IFS similarly fails to capture the strong link with Kara sea ice. Neither ERA20C nor the IFS show any notable link between the Polar vortex or the QBO with Northern persistence. The eddy feedback component (rightmost entry of the plot) will be discussed next section.

In the Southern+Central regime, the story is a little different. All the predictors seem to contribute similar levels of predictability, with ERA20C having a correlation of around $0.3$ with Kara sea-ice, ENSO and both stratospheric indices. Unlike in the Northern case, the IFS ensemble mean has comparable correlations with all of these except ENSO. This implies that while ENSO may be a driver of variability in ERA20C, it is playing little or no role in the NAO predictability of the IFS. Instead, Kara sea-ice and the stratosphere seem to be playing more important roles, with the ensemble mean correctly replicating the relationship seen for ERA20C. As before, the eddy feedback term is discussed next section.

This supports the conclusions of earlier studies, namely that Kara sea-ice and the stratosphere are playing a significant role in driving NAO predictability. Our framework suggests that these may act primarily by influencing the extent of jet persistence, particularly in the mid-latitudes. There is evidence that ENSO is playing a significant role in regulating persistence in ERA20C, especially (and curiously) in the Northern regime, but this is not replicated by the IFS.

Note also that the correlations in the two regimes are not perfectly inverse to each other. While the influence of Kara sea-ice and ENSO do naturally anti-correlate across the two (e.g. increased Kara sea-ice increases Northern persistence and simultaneously decreases Southern+Central persistence), this is not the case for the two stratospheric indices. Instead, the stratosphere seems to only modulate persistence in the Southern+Central regime, with no notable effect in the north. This again lends support to the idea that the two regimes are, to an extent, modulated independently of each other. We observe that this provides a potential quantitative way to estimate non-linear influences of teleconnections on the NAO, using the fact that the DJF mean NAO is a non-linear function of both persistence probabilities. 

Finally, we briefly address the question of dependence amongst the different predictors. This is to some extent mitigated by the choice of which time-period our predictors are considered over. In particular, our ENSO index, when restricted to November months, has almost zero correlation with both November Kara sea-ice, the November polar vortex or the DJF mean QBO index. As such, these November-only predictors are independent of the November-December-January ENSO index used. Studies have also suggested a link between Kara sea-ice and the stratospheric circulation \citep{Kim2014, Cohen2013} and indeed we find positive correlations between November sea-ice and the November polar vortex and DJF mean QBO. However, if we subtract away the sea-ice signal from the stratospheric indices using linear regression, their correlations with persistence probabilities do not appreciably change. Therefore the predictors considered appear to be broadly independent, insofar as their impact on jet persistence is considered.

\subsection{Local Eddy Feedback}
\label{sec:eddy_feedback}

\begin{figure}[p]
	\centering
    \includegraphics[width=0.8\textwidth]{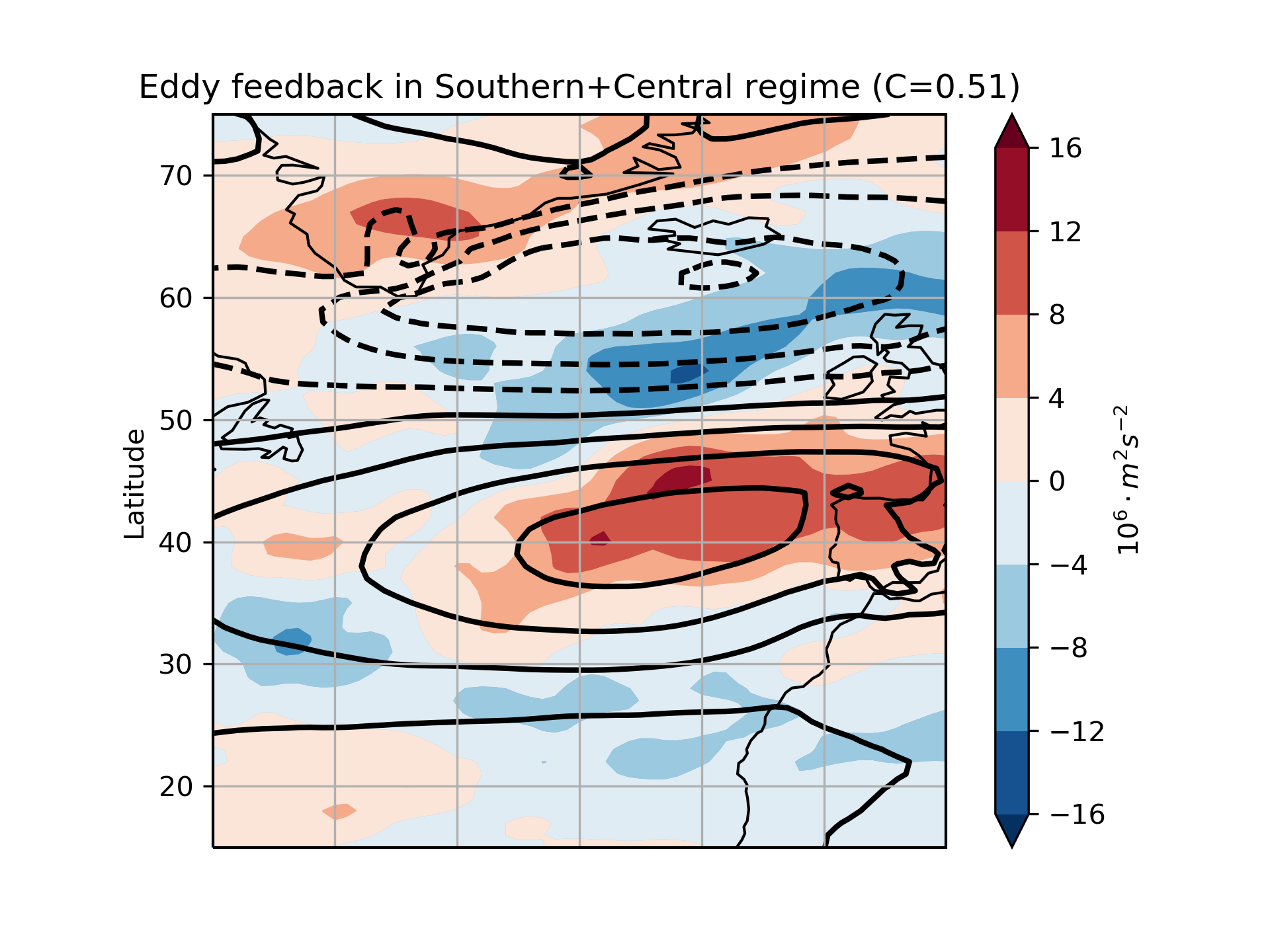}
	\caption{The filled red-blue contour shows the composite of eddy momentum flux convergence at 250hPa for days when the jet is classified as being in the Southern+Central regime (i.e. at latitudes less than 52N) minus the mean eddy momentum flux. Overlaid in black contour lines are the same composited differences for zonal 850hPa winds: solid lines are positive, dashed lines are negative. The value of $C$ in the title is the pattern correlation between the two contours. The dataset used is ERA20C and the time-period considered is 1980-2010.}
	\label{fig:eddy_contour}
\end{figure}

There is by now an extensive literature on the role of transient baroclinic eddy feedback in maintaining jet persistence (see e.g. \citet{Hoskins1983, Shutts1983, Robinson1996, Lorenz2001, Lorenz2003, Zurita-Gotor2014, Novak2015} and references therein) as well as for the NAO more directly \citep{Barnes2010}. It is therefore natural to consider the extent to which eddy feedbacks are modulating the persistence probabilities central to the Markov model. To do so we compute the daily eddy momentum flux convergence of 250hPa ERA20C winds:
\begin{equation}
E_{250} := \frac{\partial(-u_{250}''v_{250}'')}{\partial y} \nonumber
\end{equation}
where $u_{250}''$ (resp. $v_{250}''$) is the zonal (resp. meridional), 2-6 day bandpass filtered wind field at 250hPa. Positive values of this quantity correspond to regions where the transient eddies are accelerating the westerlies \citep{Hoskins1983}. \citet{Lorenz2003} showed that the eddies generally act to reinforce the jet through a positive feedback, but as noted in \citet{Woollings2010b}, this feedback is active at all latitudes, independently of the three jet regimes. The tendency of eddies to reinforce the mean flow is shown concretely in Figure \ref{fig:eddy_contour}, which shows composites of the eddy flux convergence (red-blue filled contour) and the zonal wind anomaly (overlaid black contour lines) on days when the jet is classified as being in the Southern+Central regime (that is, at latitudes lower than 52N). There is notable overlap between the two, with a positive anomaly in the region southwards of 52N, as expected. By construction, the corresponding plot for days when the jet is in the Northern regime is the same but with a sign change.

This is used to motivate the following metric, aimed at measuring the extent to which the eddies are reinforcing the jet in the Southern+Central regime. For a given winter season, consisting of 90 days, we take the dot product of the 850hPa wind anomalies and eddy flux convergence, restricted to the Southern+Central region only. The mean across all 90 daily dot products is then defined as the strength of the eddy feedback $F$ in the Southern+Central regime in this winter:
\begin{equation}
F = [E_{250} \cdot u_{850}'], \nonumber
\end{equation}
where $u_{850}'$ is the 850hPa daily, wind-anomaly, $[ \cdot ]$ denotes the DJF mean and all vectors have been restricted to the latitudes 15N-52N.
The daily dot product will be positive on days when the eddies are strongly reinforcing the jet. Figure \ref{fig:eddy_contour} implies that this will, on average, be the case and hence $F$ would be expected to correlate with measures of seasonal jet persistence. While an analogous metric for the Northern regime could be considered, the results of \citet{Barnes2010} suggest that eddy feedback is not playing a major role in the higher latitudes. We found that a seasonal dot product based on the Northern regime alone was on average an order of magnitude smaller than the one for the Southern+Central regime, corroborating this reduced influence of eddies. For these reasons we will only consider the eddy feedback in the Southern+Central region as a possible predictor of persistence. 

The correlations that the Southern+Central eddy forcing has with the persistence probabilities in the two regimes is shown in Figure \ref{fig:predictors_correlations} (the rightmost entries), confirming our intuitive expectations. For both ERA20C and the IFS ensemble mean, the correlation in the case of the Southern+Central regime is around $0.62$, significantly higher than that of any of the external predictors considered. While all these external predictors do enjoy non-zero correlations with the eddy forcing, if we subtract the influence of all these from the eddy forcing (by performing a multiple linear regression) the correlation between the forcing the Southern+Central persistence probabilities still remains high at around $0.45$. Note that this is still larger than that of all the external predictors, implying that while some of the eddy feedback is externally forced, a considerable amount of it may be either locally driven or unforced and essentially chaotic.

In the Northern regime, the eddy forcing has a smaller, negative correlation with seasonal persistence of around $-0.3$. This is consistent with the analysis of Section \ref{sec:model_skill}, which suggested that changes in the persistence characteristics of the Northern regime are merely a response to changes in the south.

\section{Conclusions}
\label{sec:conclusions}

By truncating the trimodal jet into a bimodal system, we constructed a statistical model (the Markov model) based on capturing regime dynamics as in \citet{Strommen2019}. In this system, the position of the jet is modelled using a two-state Markov chain which is driven by seasonal persistence probabilities, with the two states representing the Northern regime and the combined Southern and Central jet regimes. These persistence probabilities, assumed constant during a given winter but which otherwise vary from year to year, determine which of the two regimes the jet is likely to spend more time in. Because the average NAO conditions differ for the two regimes, with the more southern regime being associated with negative NAO values and the more northern regime associated with positive values, such interannual variations in the preferred jet regime translates into interannual variations in the NAO. Hence any knowledge of the expected persistence time-scales of the two regimes yields predictive power for the winter NAO.

We then showed that the IFS has robust skill at predicting these seasonal persistence probabilities over the modern period 1980-2010. This justified fitting the Markov model parameters using reanalysis and IFS model data, the latter in the form of a 51-member ensemble hindcast. The fitted model allowed us to estimate what the expected IFS ensemble mean correlation with the true NAO index would be if predictability were driven entirely by regime dynamics. Since this estimate, of $0.51 \pm 0.20$, is close to the observed correlation of $0.49$, we concluded that all of the skill of the IFS may be coming purely from such regime dynamics. Because the Markov model does not include any contribution from the jet speed, this conclusion is consistent with that of \citet{Parker2019}, which used linear regression techniques to attribute the IFS skill to the jet latitude. However, while linear techniques first requires one to collapse the trimodal structure of the jet by averaging, the Markov model respects the non-linear structure inherent to the jet in a novel manner. A consequence of this is that the Markov model naturally produces a negatively skewed NAO, in accordance with observations. By highlighting the role of weak model persistence, this model also suggests a possible explanation of the `signal-to-noise paradox'. Thinking of regimes as potential wells, the weak persistence of the IFS may be a result of these potential wells being too shallow. As a result, any external forcing of the jet which attempts to push the jet preferentially towards either regime would not elicit as strong a response in the model as it ought to; the jet of the model is simply ejected from the relevant regime too swiftly.

While our statistical model captures predictability associated with transitions in the jet latitude, we omitted any contribution from the jet speed, which is likely overly simplistic given that studies suggest external drivers like ENSO can influence jet speeds \citep{Drouard2013}. In addition, earlier studies suggest that there is further intra-seasonal structure in jet variability, such as preferred transitions \citep{Franzke2011, Novak2015}, which are only indirectly represented in the Markov model. There is also evidence that thermal processes can play a crucial role in driving jet transitions \citep{Novak2015}, particularly for the southern regime \citep{Madonna2019}. As these are effectively distinct from the eddy-driven processes associated with higher levels of persistence, it is possible that this is a source of IFS jet latitude skill not represented in the Markov model.

Finally, we considered both local and external drivers of variability in jet regime persistence. We found that Kara sea-ice, ENSO, the QBO and the stratospheric polar vortex, commonly cited as important drivers of NAO variability, all correlate with the estimated persistence probabilities from reanalysis. This suggests that the role of these teleconnections may be, at least in part, to modulate jet persistence rather than the NAO more directly. In the IFS, several of these external teleconnections were weaker than in reanalysis, with some being virtually non-existent. In particular, the IFS does not replicate the observed connection between ENSO and jet persistence, suggesting that ENSO may have very limited influence on skill. Locally we showed that a measure of the strength of the transient baroclinic eddy feedback also projects strongly onto the two persistence probabilities, particularly that of the southern regime. It appears that a substantial part of this eddy forcing is independent of the external drivers we considered, implying that much of the observed NAO predictability may not be easily attributable to remote teleconnections and may rather come from more complex local dynamics. This suggests that eddy feedbacks play a crucial role in modulating NAO predictability and models that fail to produce realistic levels of feedback may struggle to achieve high NAO skill even with accurate teleconnections. In \cite{Scaife2019}, by comparing seasonal predictions at varying horizontal resolutions it was found that the strength of the eddy forcing on the NAO was systematically too weak, with no appreciable change until the resolution approached $\approx 15$km grid-boxes. The importance of horizontal resolution for representing North Atlantic regime behaviour was also demonstrated in \citet{Dawson2012} and \citet{Strommen2019a}, implying that relatively high horizontal resolution may be necessary to eliminate the signal-to-noise paradox.

We end our discussion with two questions. Firstly, the results of this paper were based on the IFS in particular. In \citet{Baker2018} it was found that models can exhibit a broad range of ensemble mean correlations and $RPC$ values: can their differing abilities at capturing regime persistence explain this variability? Secondly, in the jet regime system captured by the Markov model, predictability is explicitly induced by variations in persistence probabilities, which are strongly modulated by local transient eddy feedback. Should the role of more remote teleconnections also be understood, not through their direct action on the NAO, but their indirect influence on jet persistence via eddy feedbacks?


\acks
This work benefited from conversations with Tim Woollings, Tim Palmer, Fenwick Cooper, Marie Drouard and Peter Watson. It was carried out with funding from the European Commission under Grant Agreement 641727 of the Horizon 2020 research programme. Processed IFS and ERA20C wind data was generously shared with me by Tess Parker. Processed QBO data was generously shared with me by Chris O'Reilly. Processed Kara sea-ice data was generously shared with me by Fenwick Cooper.

\newpage
\bibliographystyle{wileyqj}           
\bibliography{jet_references}{}

\begin{thebibliography}{63}
\providecommand{\natexlab}[1]{#1}
\providecommand{\url}[1]{\texttt{#1}}
\providecommand{\urlprefix}{URL }
\expandafter\ifx\csname urlstyle\endcsname\relax
  \providecommand{\doi}[1]{doi:\discretionary{}{}{}#1}\else
  \providecommand{\doi}{doi:\discretionary{}{}{}\begingroup
  \urlstyle{rm}\Url}\fi

\bibitem[{Anstey \emph{et~al.}(2013)Anstey, Davini, Gray, Woollings, Butchart,
  Cagnazzo, Christiansen, Hardiman, Osprey and Yang}]{Anstey2013}
Anstey JA, Davini P, Gray LJ, Woollings TJ, Butchart N, Cagnazzo C,
  Christiansen B, Hardiman SC, Osprey SM, Yang S. 2013. {Multi-model analysis
  of Northern Hemisphere winter blocking: Model biases and the role of
  resolution}. \emph{Journal of Geophysical Research Atmospheres}
  \textbf{118}(10): 3956--3971, \doi{10.1002/jgrd.50231}.

\bibitem[{Athanasiadis \emph{et~al.}(2017)Athanasiadis, Bellucci, Scaife,
  Hermanson, Materia, Sanna, Borrelli, MacLachlan and
  Gualdi}]{Athanasiadis2017}
Athanasiadis PJ, Bellucci A, Scaife AA, Hermanson L, Materia S, Sanna A,
  Borrelli A, MacLachlan C, Gualdi S. 2017. {A multisystem view of wintertime
  NAO seasonal predictions}. \emph{Journal of Climate} \textbf{30}(4):
  1461--1475, \doi{10.1175/JCLI-D-16-0153.1}.

\bibitem[{Athanasiadis \emph{et~al.}(2010)Athanasiadis, Wallace and
  Wettstein}]{Athanasiadis2010}
Athanasiadis PJ, Wallace JM, Wettstein JJ. 2010. {Patterns of wintertime jet
  stream variability and their relation to the storm tracks}. \emph{Journal of
  the Atmospheric Sciences} \doi{10.1175/2009JAS3270.1}.

\bibitem[{Baker \emph{et~al.}(2018)Baker, Shaffrey, Sutton, Weisheimer and
  Scaife}]{Baker2018}
Baker LH, Shaffrey LC, Sutton RT, Weisheimer A, Scaife AA. 2018. {An
  Intercomparison of Skill and Overconfidence/Underconfidence of the Wintertime
  North Atlantic Oscillation in Multimodel Seasonal Forecasts}.
  \emph{Geophysical Research Letters} \doi{10.1029/2018GL078838}.

\bibitem[{Baldwin and Dunkerton(2001)}]{Baldwin2001}
Baldwin MP, Dunkerton TJ. 2001. {Stratospheric harbingers of anomalous weather
  regimes}. \emph{Science} \doi{10.1126/science.1063315}.

\bibitem[{Barnes and Hartmann(2010)}]{Barnes2010}
Barnes EA, Hartmann DL. 2010. {Dynamical Feedbacks and the Persistence of the
  NAO}. \emph{Journal of the Atmospheric Sciences} \doi{10.1175/2009JAS3193.1}.

\bibitem[{Br{\"{o}}nnimann(2007)}]{Bronnimann2007}
Br{\"{o}}nnimann S. 2007. {Impact of El Ni{\~{n}}o-Southern Oscillation on
  European climate}. \doi{10.1029/2006RG000199}.

\bibitem[{Charlton-Perez \emph{et~al.}(2018)Charlton-Perez, Ferranti and
  Lee}]{Charlton-Perez2018}
Charlton-Perez AJ, Ferranti L, Lee RW. 2018. {The influence of the
  stratospheric state on North Atlantic weather regimes}. \emph{Quarterly
  Journal of the Royal Meteorological Society} \doi{10.1002/qj.3280}.

\bibitem[{Cohen \emph{et~al.}(2013)Cohen, Jones, Furtado and
  Tziperman}]{Cohen2013}
Cohen J, Jones J, Furtado JC, Tziperman E. 2013. {Warm Arctic, cold continents
  a common pattern related to Arctic sea ice melt, snow advance, and extreme
  winter weather}. \emph{Oceanography} \doi{10.5670/oceanog.2013.70}.

\bibitem[{Czaja and Frankignoul(1999)}]{Czaja1999}
Czaja A, Frankignoul C. 1999. {Influence of the North Atlantic SST on the
  atmospheric circulation}. \emph{Geophysical Research Letters}
  \doi{10.1029/1999GL900613}.

\bibitem[{Dawson \emph{et~al.}(2012)Dawson, Palmer and Corti}]{Dawson2012}
Dawson A, Palmer TN, Corti S. 2012. {Simulating regime structures in weather
  and climate prediction models}. \emph{Geophysical Research Letters}
  \textbf{39}(21), \doi{10.1029/2012GL053284}.

\bibitem[{Dee \emph{et~al.}(2011)Dee, Uppala, Simmons, Berrisford, Poli,
  Kobayashi, Andrae, Balmaseda, Balsamo, Bauer, Bechtold, Beljaars, van~de
  Berg, Bidlot, Bormann, Delsol, Dragani, Fuentes, Geer, Haimberger, Healy,
  Hersbach, H{\'{o}}lm, Isaksen, K{\aa}llberg, K{\"{o}}hler, Matricardi,
  Mcnally, Monge-Sanz, Morcrette, Park, Peubey, de~Rosnay, Tavolato,
  Th{\'{e}}paut and Vitart}]{Dee2011}
Dee DP, Uppala SM, Simmons AJ, Berrisford P, Poli P, Kobayashi S, Andrae U,
  Balmaseda MA, Balsamo G, Bauer P, Bechtold P, Beljaars AC, van~de Berg L,
  Bidlot J, Bormann N, Delsol C, Dragani R, Fuentes M, Geer AJ, Haimberger L,
  Healy SB, Hersbach H, H{\'{o}}lm EV, Isaksen L, K{\aa}llberg P, K{\"{o}}hler
  M, Matricardi M, Mcnally AP, Monge-Sanz BM, Morcrette JJ, Park BK, Peubey C,
  de~Rosnay P, Tavolato C, Th{\'{e}}paut JN, Vitart F. 2011. {The ERA-Interim
  reanalysis: Configuration and performance of the data assimilation system}.
  \emph{Quarterly Journal of the Royal Meteorological Society}
  \textbf{137}(656): 553--597, \doi{10.1002/qj.828}.

\bibitem[{Dong \emph{et~al.}(2000)Dong, Sutton, Jewson, O'Neill and
  Slingo}]{Dong2000}
Dong BW, Sutton RT, Jewson SP, O'Neill A, Slingo JM. 2000. {Predictable winter
  climate in the North Atlantic sector during the 1997-1999 ENSO cycle}.
  \emph{Geophysical Research Letters} \doi{10.1029/1999GL010994}.

\bibitem[{Drouard \emph{et~al.}(2013)Drouard, Rivi{\`{e}}re and
  Arbogast}]{Drouard2013}
Drouard M, Rivi{\`{e}}re G, Arbogast P. 2013. {The north Atlantic oscillation
  response to large-scale atmospheric anomalies in the northeastern Pacific}.
  \emph{Journal of the Atmospheric Sciences} \doi{10.1175/JAS-D-12-0351.1}.

\bibitem[{Dunstone \emph{et~al.}(2016)Dunstone, Smith, Scaife, Hermanson, Eade,
  Robinson, Andrews and Knight}]{Dunstone2016}
Dunstone N, Smith D, Scaife A, Hermanson L, Eade R, Robinson N, Andrews M,
  Knight J. 2016. {Skilful predictions of the winter North Atlantic Oscillation
  one year ahead}. \emph{Nature Geoscience} \textbf{9}(11): 809--814,
  \doi{10.1038/ngeo2824}.

\bibitem[{Eade \emph{et~al.}(2014)Eade, Smith, Scaife, Wallace, Dunstone,
  Hermanson and Robinson}]{Eade2014}
Eade R, Smith D, Scaife A, Wallace E, Dunstone N, Hermanson L, Robinson N.
  2014. {Do seasonal-to-decadal climate predictions underestimate the
  predictability of the real world?} \emph{Geophysical Research Letters}
  \textbf{41}(15): 5620--5628, \doi{10.1002/2014GL061146}.

\bibitem[{Franzke \emph{et~al.}(2011)Franzke, Woollings and
  Martius}]{Franzke2011}
Franzke C, Woollings T, Martius O. 2011. {Persistent Circulation Regimes and
  Preferred Regime Transitions in the North Atlantic}. \emph{Journal of the
  Atmospheric Sciences} \textbf{68}(12): 2809--2825,
  \doi{10.1175/JAS-D-11-046.1},
  \urlprefix\url{http://journals.ametsoc.org/doi/abs/10.1175/JAS-D-11-046.1}.

\bibitem[{Gastineau \emph{et~al.}(2013)Gastineau, D'Andrea and
  Frankignoul}]{Gastineau2013}
Gastineau G, D'Andrea F, Frankignoul C. 2013. {Atmospheric response to the
  North Atlantic Ocean variability on seasonal to decadal time scales}.
  \emph{Climate Dynamics} \doi{10.1007/s00382-012-1333-0}.

\bibitem[{Hoskins \emph{et~al.}(1983)Hoskins, James and White}]{Hoskins1983}
Hoskins BJ, James IN, White GH. 1983. {The shape, propagation and mean-flow
  interaction of large-scale weather systems.} \emph{Journal of the Atmospheric
  Sciences} \doi{10.1175/1520-0469(1983)040<1595:TSPAMF>2.0.CO;2}.

\bibitem[{Jim{\'{e}}nez-Esteve and Domeisen(2018)}]{Jimenez-Esteve2018}
Jim{\'{e}}nez-Esteve B, Domeisen DI. 2018. {The tropospheric pathway of the
  ENSO-North Atlantic teleconnection}. \emph{Journal of Climate}
  \doi{10.1175/JCLI-D-17-0716.1}.

\bibitem[{Kidston \emph{et~al.}(2015)Kidston, Scaife, Hardiman, Mitchell,
  Butchart, Baldwin and Gray}]{Kidston2015}
Kidston J, Scaife AA, Hardiman SC, Mitchell DM, Butchart N, Baldwin MP, Gray
  LJ. 2015. {Stratospheric influence on tropospheric jet streams, storm tracks
  and surface weather}. \emph{Nature Geoscience} \doi{10.1038/NGEO2424}.

\bibitem[{Kim \emph{et~al.}(2014)Kim, Son, Min, Jeong, Kim, Zhang, Shim and
  Yoon}]{Kim2014}
Kim BM, Son SW, Min SK, Jeong JH, Kim SJ, Zhang X, Shim T, Yoon JH. 2014.
  {Weakening of the stratospheric polar vortex by Arctic sea-ice loss}.
  \emph{Nature Communications} \doi{10.1038/ncomms5646}.

\bibitem[{Koenigk \emph{et~al.}(2016)Koenigk, Caian, Nikulin and
  Schimanke}]{Koenigk2016}
Koenigk T, Caian M, Nikulin G, Schimanke S. 2016. {Regional Arctic sea ice
  variations as predictor for winter climate conditions}. \emph{Climate
  Dynamics} \doi{10.1007/s00382-015-2586-1}.

\bibitem[{Kumar \emph{et~al.}(2014)Kumar, Peng and Chen}]{Kumar2014}
Kumar A, Peng P, Chen M. 2014. {Is There a Relationship between Potential and
  Actual Skill?} \emph{Monthly Weather Review} \textbf{142}(6): 2220--2227,
  \doi{10.1175/MWR-D-13-00287.1},
  \urlprefix\url{http://journals.ametsoc.org/doi/abs/10.1175/MWR-D-13-00287.1}.

\bibitem[{Li and Lau(2012)}]{Li2012}
Li Y, Lau NC. 2012. {Impact of ENSO on the atmospheric variability over the
  North Atlantic in late Winter-Role of transient eddies}. \emph{Journal of
  Climate} \doi{10.1175/JCLI-D-11-00037.1}.

\bibitem[{Lorenz and Hartmann(2001)}]{Lorenz2001}
Lorenz DJ, Hartmann DL. 2001. {Eddy-zonal flow feedback in the Southern
  Hemisphere}. \emph{Journal of the Atmospheric Sciences}
  \doi{10.1175/1520-0469(2001)058<3312:EZFFIT>2.0.CO;2}.

\bibitem[{Lorenz and Hartmann(2003)}]{Lorenz2003}
Lorenz DJ, Hartmann DL. 2003. {Eddy-zonal flow feedback in the Northern
  Hemisphere winter}. \emph{Journal of Climate}
  \doi{10.1175/1520-0442(2003)16<1212:EFFITN>2.0.CO;2}.

\bibitem[{Madonna \emph{et~al.}(2019)Madonna, Li and Wettstein}]{Madonna2019}
Madonna E, Li C, Wettstein JJ. 2019. Suppressed eddy driving during southward
  excursions of the north atlantic jet on synoptic to seasonal time scales.
  \emph{Atmospheric Science Letters} \textbf{20}(9): e937,
  \doi{10.1002/asl.937},
  \urlprefix\url{https://rmets.onlinelibrary.wiley.com/doi/abs/10.1002/asl.937}.

\bibitem[{Mehta \emph{et~al.}(2000)Mehta, Suarez, Manganello and
  Delworth}]{Mehta2000}
Mehta VM, Suarez MJ, Manganello JV, Delworth TL. 2000. {Oceanic influence on
  the North Atlantic Oscillation and associated Northern Hemisphere climate
  variations: 1959-1993}. \emph{Geophysical Research Letters}
  \doi{10.1029/1999GL002381}.

\bibitem[{Michelangeli \emph{et~al.}(1995)Michelangeli, Vautard and
  Legras}]{Michelangeli1995a}
Michelangeli PA, Vautard R, Legras B. 1995. {Weather Regimes: Recurrence and
  Quasi Stationarity}. \emph{Journal of the Atmospheric Sciences}
  \textbf{52}(8): 1237--1256,
  \doi{10.1175/1520-0469(1995)052<1237:WRRAQS>2.0.CO;2},
  \urlprefix\url{http://journals.ametsoc.org/doi/abs/10.1175/1520-0469{\%}281995{\%}29052{\%}3C1237{\%}3AWRRAQS{\%}3E2.0.CO{\%}3B2}.

\bibitem[{Moron and Gouirand(2003)}]{Moron2003}
Moron V, Gouirand I. 2003. {Seasonal modulation of the El Ni{\~{n}}o-southern
  oscillation relationship with sea level pressure anomalies over the North
  Atlantic in October-March 1873-1996}. \emph{International Journal of
  Climatology} \doi{10.1002/joc.868}.

\bibitem[{M{\"{u}}ller \emph{et~al.}(2005)M{\"{u}}ller, Appenzeller and
  Sch{\"{a}}r}]{Muller2005}
M{\"{u}}ller WA, Appenzeller C, Sch{\"{a}}r C. 2005. {Probabilistic seasonal
  prediction of the winter North Atlantic Oscillation and its impact on near
  surface temperature}. \emph{Climate Dynamics} \textbf{24}(2-3): 213--226,
  \doi{10.1007/s00382-004-0492-z}.

\bibitem[{Novak \emph{et~al.}(2015)Novak, Ambaum and Tailleux}]{Novak2015}
Novak L, Ambaum MH, Tailleux R. 2015. {The life cycle of the North Atlantic
  storm track}. \emph{Journal of the Atmospheric Sciences}
  \doi{10.1175/JAS-D-14-0082.1}.

\bibitem[{O'Reilly \emph{et~al.}(2017)O'Reilly, Heatley, MacLeod, Weisheimer,
  Palmer, Schaller and Woollings}]{OReilly2017}
O'Reilly CH, Heatley J, MacLeod D, Weisheimer A, Palmer TN, Schaller N,
  Woollings T. 2017. {Variability in seasonal forecast skill of Northern
  Hemisphere winters over the twentieth century}. \emph{Geophysical Research
  Letters} \doi{10.1002/2017GL073736}.

\bibitem[{O'Reilly \emph{et~al.}(2019)O'Reilly, Weisheimer, Woollings, Gray and
  MacLeod}]{OReilly2019}
O'Reilly CH, Weisheimer A, Woollings T, Gray LJ, MacLeod D. 2019. {The
  importance of stratospheric initial conditions for winter North Atlantic
  Oscillation predictability and implications for the signal-to-noise paradox}.
  \emph{Quarterly Journal of the Royal Meteorological Society}
  \doi{10.1002/qj.3413}.

\bibitem[{Palmer(1999)}]{Palmer1999}
Palmer TN. 1999. {A nonlinear dynamical perspective on climate prediction}.
  \emph{Journal of Climate} \textbf{12}(2): 575--591,
  \doi{10.1002/j.1477-8696.1993.tb05802.x}.

\bibitem[{Parker \emph{et~al.}(2019)Parker, Woollings, Weisheimer, O'Reilly,
  Baker and Shaffrey}]{Parker2019}
Parker T, Woollings T, Weisheimer A, O'Reilly C, Baker L, Shaffrey L. 2019.
  Seasonal predictability of the winter north atlantic oscillation from a jet
  stream perspective. \emph{Geophysical Research Letters} \textbf{0}(0),
  \doi{10.1029/2019GL084402},
  \urlprefix\url{https://agupubs.onlinelibrary.wiley.com/doi/abs/10.1029/2019GL084402}.

\bibitem[{Poli \emph{et~al.}(2013)Poli, Hersbach, Tan, Dee, Th{\'{e}}paut,
  Simmons, Peubey, Laloyaux, Komori, Berrisford and Dragani}]{Poli2013}
Poli P, Hersbach H, Tan D, Dee D, Th{\'{e}}paut JN, Simmons A, Peubey C,
  Laloyaux P, Komori T, Berrisford P, Dragani R. 2013. {The data assimilation
  system and initial performance evaluation of the ECMWF pilot reanalysis of
  the 20th-century assimilating surface observations only (ERA-20C)}. \emph{ERA
  report series} .

\bibitem[{Rayner \emph{et~al.}(2003)Rayner, Parker, Horton, Folland, Alexander,
  Rowell, Kent and Kaplan}]{Rayner2003}
Rayner NA, Parker DE, Horton EB, Folland CK, Alexander LV, Rowell DP, Kent EC,
  Kaplan A. 2003. {Global analyses of sea surface temperature, sea ice, and
  night marine air temperature since the late nineteenth century}.
  \emph{Journal of Geophysical Research D: Atmospheres} .

\bibitem[{Robinson(1996)}]{Robinson1996}
Robinson WA. 1996. {Does eddy feedback sustain variability in the zonal index?}
  \emph{Journal of the Atmospheric Sciences}
  \doi{10.1175/1520-0469(1996)053<3556:DEFSVI>2.0.CO;2}.

\bibitem[{Rodwell \emph{et~al.}(1999)Rodwell, Rowell and Folland}]{Rodwell1999}
Rodwell MJ, Rowell DP, Folland CK. 1999. {Oceanic forcing of the wintertime
  North Atlantic Oscillation and European climate}. \emph{Nature}
  \doi{10.1038/18648}.

\bibitem[{Rogers(1997)}]{Rogers1997}
Rogers JC. 1997. {North Atlantic storm track variability and its association to
  the North Atlantic oscillation and climate variability of Northern Europe}.
  \emph{Journal of Climate}
  \doi{10.1175/1520-0442(1997)010<1635:NASTVA>2.0.CO;2}.

\bibitem[{Saunders and Qian(2002)}]{Saunders2002}
Saunders MA, Qian B. 2002. {Seasonal predictability of the winter NAO from
  north Atlantic sea surface temperatures}. \emph{Geophysical Research Letters}
  \doi{10.1029/2002gl014952}.

\bibitem[{Scaife \emph{et~al.}(2014{\natexlab{a}})Scaife, Arribas, Blockley,
  Brookshaw, Clark, Dunstone, Eade, Fereday, Folland, Gordon, Hermanson,
  Knight, Lea, MacLachlan, Maidens, Martin, Peterson, Smith, Vellinga, Wallace,
  Waters and Williams}]{Scaife2014a}
Scaife AA, Arribas A, Blockley E, Brookshaw A, Clark RT, Dunstone N, Eade R,
  Fereday D, Folland CK, Gordon M, Hermanson L, Knight JR, Lea DJ, MacLachlan
  C, Maidens A, Martin M, Peterson AK, Smith D, Vellinga M, Wallace E, Waters
  J, Williams A. 2014{\natexlab{a}}. {Skillful long-range prediction of
  European and North American winters}. \emph{Geophysical Research Letters}
  \textbf{41}(7): 2514--2519, \doi{10.1002/2014GL059637}.

\bibitem[{Scaife \emph{et~al.}(2014{\natexlab{b}})Scaife, Athanassiadou,
  Andrews, Arribas, Baldwin, Dunstone, Knight, Maclachlan, Manzini,
  M{\"{u}}ller, Pohlmann, Smith, Stockdale and Williams}]{Scaife2014}
Scaife AA, Athanassiadou M, Andrews M, Arribas A, Baldwin M, Dunstone N, Knight
  J, Maclachlan C, Manzini E, M{\"{u}}ller WA, Pohlmann H, Smith D, Stockdale
  T, Williams A. 2014{\natexlab{b}}. {Predictability of the quasi-biennial
  oscillation and its northern winter teleconnection on seasonal to decadal
  timescales}. \emph{Geophysical Research Letters} \doi{10.1002/2013GL059160}.

\bibitem[{Scaife \emph{et~al.}(2019)Scaife, Camp, Comer, Davis, Dunstone,
  Gordon, MacLachlan, Martin, Nie, Ren, Roberts, Robinson, Smith and
  Vidale}]{Scaife2019}
Scaife AA, Camp J, Comer R, Davis P, Dunstone N, Gordon M, MacLachlan C, Martin
  N, Nie Y, Ren HL, Roberts M, Robinson W, Smith D, Vidale PL. 2019. {Does
  increased atmospheric resolution improve seasonal climate predictions?}
  \emph{Atmospheric Science Letters} \doi{10.1002/asl.922}.

\bibitem[{Scaife \emph{et~al.}(2013)Scaife, Ineson, Knight, Gray, Kodera and
  Smith}]{Scaife2013}
Scaife AA, Ineson S, Knight JR, Gray L, Kodera K, Smith DM. 2013. {A mechanism
  for lagged North Atlantic climate response to solar variability}.
  \emph{Geophysical Research Letters} \doi{10.1002/grl.50099}.

\bibitem[{Scaife and Smith(2018)}]{Scaife2018}
Scaife AA, Smith D. 2018. {A signal-to-noise paradox in climate science}.
  \emph{npj Climate and Atmospheric Science} \doi{10.1038/s41612-018-0038-4}.

\bibitem[{Shutts(1983)}]{Shutts1983}
Shutts GJ. 1983. {The propagation of eddies in diffluent jetstreams: Eddy
  vorticity forcing of ‘blocking' flow fields}. \emph{Quarterly Journal of
  the Royal Meteorological Society} \doi{10.1002/qj.49710946204}.

\bibitem[{Siegert \emph{et~al.}(2016)Siegert, Stephenson, Sansom, Scaife, Eade
  and Arribas}]{Siegert2016}
Siegert S, Stephenson DB, Sansom PG, Scaife AA, Eade R, Arribas A. 2016. {A
  Bayesian framework for verification and recalibration of ensemble forecasts:
  How uncertain is NAO predictability?} \emph{Journal of Climate}
  \textbf{29}(3): 995--1012, \doi{10.1175/JCLI-D-15-0196.1}.

\bibitem[{Smith \emph{et~al.}(2016)Smith, Scaife, Eade and Knight}]{Smith2016}
Smith DM, Scaife AA, Eade R, Knight JR. 2016. {Seasonal to decadal prediction
  of the winter North Atlantic Oscillation: Emerging capability and future
  prospects}. \emph{Quarterly Journal of the Royal Meteorological Society}
  \textbf{142}(695): 611--617, \doi{10.1002/qj.2479}.

\bibitem[{Strommen \emph{et~al.}(2019)Strommen, Mavilia, Corti, Matsueda,
  Davini, von Hardenberg, Vidale and Mizuta}]{Strommen2019a}
Strommen K, Mavilia I, Corti S, Matsueda M, Davini P, von Hardenberg J, Vidale
  PL, Mizuta R. 2019. {The Sensitivity of Euro-Atlantic Regimes to Model
  Horizontal Resolution}. \emph{Geophysical Research Letters}
  \doi{10.1029/2019GL082843}.

\bibitem[{Strommen and Palmer(2019)}]{Strommen2019}
Strommen K, Palmer TN. 2019. {Signal and noise in regime systems: A hypothesis
  on the predictability of the North Atlantic Oscillation}. \emph{Quarterly
  Journal of the Royal Meteorological Society} \doi{10.1002/qj.3414}.

\bibitem[{Strong and Magnusdottir(2011)}]{Strong2011}
Strong C, Magnusdottir G. 2011. {Dependence of NAO variability on coupling with
  sea ice}. \emph{Climate Dynamics} \doi{10.1007/s00382-010-0752-z}.

\bibitem[{Thompson \emph{et~al.}(2003)Thompson, Lee and Baldwin}]{Thompson2003}
Thompson DW, Lee S, Baldwin MP. 2003. {Atmospheric processes governing the
  northern hemisphere annular mode/north atlantic oscillation}. In:
  \emph{Geophysical Monograph Series}, ISBN 9781118669037,
  \doi{10.1029/134GM05}.

\bibitem[{Toniazzo and Scaife(2006)}]{Toniazzo2006}
Toniazzo T, Scaife AA. 2006. {The influence of ENSO on winter North Atlantic
  climate}. \emph{Geophysical Research Letters} \doi{10.1029/2006GL027881}.

\bibitem[{Vautard and Vautard(1990)}]{Vautard1990}
Vautard R, Vautard R. 1990. {Multiple Weather Regimes over the North Atlantic:
  Analysis of Precursors and Successors}. \emph{Monthly Weather Review}
  \doi{10.1175/1520-0493(1990)118<2056:MWROTN>2.0.CO;2}.

\bibitem[{Wang \emph{et~al.}(2017)Wang, Ting and Kushner}]{Wang2017}
Wang L, Ting M, Kushner PJ. 2017. {A robust empirical seasonal prediction of
  winter NAO and surface climate}. \emph{Scientific Reports}
  \doi{10.1038/s41598-017-00353-y}.

\bibitem[{Weisheimer \emph{et~al.}(2017)Weisheimer, Schaller, O'Reilly, MacLeod
  and Palmer}]{Weisheimer2017}
Weisheimer A, Schaller N, O'Reilly C, MacLeod DA, Palmer T. 2017. {Atmospheric
  seasonal forecasts of the twentieth century: multi-decadal variability in
  predictive skill of the winter North Atlantic Oscillation (NAO) and their
  potential value for extreme event attribution}. \emph{Quarterly Journal of
  the Royal Meteorological Society} \doi{10.1002/qj.2976}.

\bibitem[{Woollings \emph{et~al.}(2010{\natexlab{a}})Woollings, Charlton-Perez,
  Ineson, Marshall and Masato}]{Woollings2010}
Woollings T, Charlton-Perez A, Ineson S, Marshall AG, Masato G.
  2010{\natexlab{a}}. {Associations between stratospheric variability and
  tropospheric blocking}. \emph{Journal of Geophysical Research Atmospheres}
  \textbf{115}(6), \doi{10.1029/2009JD012742}.

\bibitem[{Woollings \emph{et~al.}(2010{\natexlab{b}})Woollings, Hannachi and
  Hoskins}]{Woollings2010b}
Woollings T, Hannachi A, Hoskins B. 2010{\natexlab{b}}. {Variability of the
  North Atlantic eddy-driven jet stream}. \emph{Quarterly Journal of the Royal
  Meteorological Society} \textbf{136}(649): 856--868, \doi{10.1002/qj.625}.

\bibitem[{Woollings \emph{et~al.}(2008)Woollings, Hoskins, Blackburn and
  Berrisford}]{Woollings2008}
Woollings T, Hoskins B, Blackburn M, Berrisford P. 2008. {A New Rossby
  Wave-Breaking Interpretation of the North Atlantic Oscillation}.
  \emph{Journal of the Atmospheric Sciences} \textbf{65}(2): 609--626,
  \doi{10.1175/2007JAS2347.1},
  \urlprefix\url{http://journals.ametsoc.org/doi/abs/10.1175/2007JAS2347.1}.

\bibitem[{Zurita-Gotor \emph{et~al.}(2014)Zurita-Gotor, Blanco-Fuentes and
  Gerber}]{Zurita-Gotor2014}
Zurita-Gotor P, Blanco-Fuentes J, Gerber EP. 2014. {The impact of baroclinic
  eddy feedback on the persistence of jet variability in the two-layer model}.
  \emph{Journal of the Atmospheric Sciences} \doi{10.1175/JAS-D-13-0102.1}.

\end{thebibliography}

\newpage
\section{Appendix}
\label{appendixA}

We describe how the fitted parameters of the Markov model were obtained.

The basic tool used is the simple `moment estimator' method described in \citet{Siegert2016}. This method uses the explicit equations to compute theoretical formula for various statistical quantities. By equating these to estimates of these quantities obtained using data one gets estimates of the various model parameters. This is described in Section 2 and Appendix B of ibid. For completeness, we sketch the method here as well.

Assume one has an arbitrary `signal+noise' system as follows:
\begin{eqnarray}
y &=& s + \epsilon + \mu_y \nonumber \\
x &=& \beta s + \eta + \mu_x \nonumber
\end{eqnarray}
where $y$ is the `true' value to be predicted and $x$ the model prediction. Here $s \sim \mathcal{N}(0,\sigma_s), \epsilon \sim \mathcal{N}(0, \sigma_{\epsilon}), \eta \sim \mathcal{N}(0, \sigma_{\eta})$ are all normally distributed. The parameters $\mu_y$ and $\mu_x$ are constants representing the means of $y$ and $x$, with $\beta$ a constant parameter capturing model skill. It is assumed that the data available for fitting are long time-series of $y=y_t$ and $x=x_{t,r}$, for ensemble members $r=1, \ldots, R$ and time-steps $t=1, \ldots, N$. Then the moment estimator method estimates values of the parameters in order as follows:
\begin{itemize}
\item[(1)] The values of $\mu_x$ and $\mu_y$ are equated with the means of $x$ and $y$ respectively.
\item[(2)] The value of $\sigma_{\eta}^2$ is equated with the average ensemble variance $v_x$:
\begin{equation}
v_x = (NR)^{-1} \sum_{t=1}^{N} \sum_{r=1}^R (x_{t,r} - \overline{x}_t)^2 \nonumber
\end{equation}
with $\overline{x}_t$ representing the ensemble mean.
\item[(3)] The value of $\beta$ is estimated as
\begin{equation}
\beta = \frac{1}{Cov(\overline{x}_t, y)} \cdot (Var(\overline{x}_t) - R^{-1}v_x). \nonumber
\end{equation}
\item[(4)] The value of $\sigma_s$ is determined via
\begin{equation}
\sigma_s^2 = \beta^{-1} Cov(\overline{x}_t, y). \nonumber
\end{equation}
\item[(5)] Finally, $\sigma_{\epsilon}$ is determined via
\begin{equation}
\sigma_{\epsilon}^2 = Var(y) - \sigma_s^2. \nonumber
\end{equation}
\end{itemize}
This gives an algorithm for estimating parameters given an arbitrary such system. This is used to fit parameters to the system of equations (7)-(10) in Section \ref{sec:model_skill} as follows.

First apply the algorithm to the signal+noise system defined by equations (7) and (8), using the estimated persistence probabilities from ERA20C and the IFS ensemble. This gives estimates of all the parameters used in those two equations. Next, consider the signal+noise system defined by equations (7) and (10). Step 3 of the algorithm gives an explicit estimate of the quantity $\beta_B \lambda \approx -0.36$. Next, note that
\begin{equation}
Cov(\hat{p}_{\text{obs}}, \hat{q}_{\text{obs}}) = \lambda Var(s). \nonumber
\end{equation}
Computing the covariance and substituting in the estimate of $Var(s)$ obtained already implies that $\lambda \approx -0.95$ and hence $\beta_B \approx 0.37$. The other parameters of equations (9) and (10) are now obtained by applying steps 1, 2, 4 and 5 of the algorithm. This produces, in totality, the numbers listed in Table \ref{tab:parameter_fits}.
\newpage


\begin{table}[h] 
\caption{Estimated parameters for the signal+noise decompositions of persistence probabilities.}
\label{tab:parameter_fits} 
\begin{tabular}{l*{2}{c}r}
\hline\noalign{\smallskip}
Parameter & Estimate \\
\noalign{\smallskip}\hline\noalign{\smallskip}
$\sigma_s$ & 0.46 \\
$\sigma_{\epsilon_A}$ & 0.55 \\
$\beta_A$ & 0.25 \\
$\sigma_{\eta_A}$ & 0.62 \\
$\sigma_{\epsilon_B}$ & 0.56 \\
$\sigma_{\eta_B}$ & 0.68 \\
$\beta_B$ & 0.37 \\
$\lambda$ & -0.95 \\
$\mu_{\text{obs}, A}$ & 2.72 \\
$\mu_{\text{obs}, B}$ & 2.11 \\
$\mu_{\text{mod}, A}$ & 2.70 \\
$\mu_{\text{mod}, B}$ & 1.81 \\
\noalign{\smallskip}\hline
\end{tabular}
\end{table}
\clearpage

\end{document}